\documentclass[12pt]{iopart}
\usepackage{graphicx}
\usepackage{wrapfig}
\usepackage{caption}
\usepackage{subcaption}
\usepackage{amssymb, bm}
\usepackage{color}
\usepackage[colorlinks,
            linkcolor=blue,
            anchorcolor=blue,
            citecolor=green
            ]{hyperref}
\usepackage{booktabs}
\usepackage{ulem}
\usepackage{xcolor}
\graphicspath{{Figures/}}

\def\eos#1{equation of state#1 (EOS#1)\gdef\eos{EOS}}
\def\QNM#1{quasi-normal mode#1 (QNM#1)\gdef\QNM{QNM}}
\def\ns#1{neutron star#1 (NS#1)\gdef\ns{NS}}
\def\gw#1{gravitational wave#1 (GW#1)\gdef\gw{GW}}
\def\bh#1{black hole#1 (BH#1)\gdef\bh{BH}}
\def\bbh#1{binary black hole#1  (BBH#1)\gdef\bbh{BBH}}
\def\bns#1{binary neutron star#1 (BNS#1)\gdef\bns{BHS}}
\def\bhns#1{black hole - neutron star#1 (BHNS#1)\gdef\bhns{BHNS}}
\def\nsbh#1{neutron star - black hole#1 (NSBH#1)\gdef\nsbh{NSBH}}
\def\nr#1{numerical relativity#1 (NR#1)\gdef\nr{NR}}
\def\ligo#1{Laser Interferometer Gravitational-wave Observatory#1 (LIGO#1)\gdef\ligo{LIGO}}

\newcommand{\tabhead}[1]{\textbf{#1}}

\graphicspath{ {Figures/}}

\begin{document}

\title{Black Hole - Neutron Star Binary Mergers: The Impact of Stellar Compactness}

\author{
    Bing-Jyun Tsao,
    Bhavesh Khamesra,
    Miguel Gracia-Linares,
    Pablo Laguna
}
\address
{Center for Gravitational Physics, Department of Physics, The University of Texas at Austin, Austin, TX 78712, U.S.A.
} 

\begin{abstract}
Recent gravitational wave observations include possible detections of black hole - neutron star binary mergers. As with binary black hole mergers, numerical simulations help characterize the sources. For binary systems with neutron star components, the simulations help to predict the imprint  of tidal deformations and disruptions on the gravitational wave signals. In a previous study, we investigated how the mass of the black hole has an impact on the disruption of the neutron star and, as a consequence, on the shape of the gravitational waves emitted. We extend these results to study the effects of varying the compactness of the neutron star. We consider neutron star compactness in the 0.123 to 0.2 range for binaries with mass ratios of 3 and 5. As the compactness and the mass ratio increase, the binary system behaves during the late inspiral and merger more like a black hole binary.  For the case with the highest mass ratio and most compact neutron star, the gravitational waves emitted, in terms of mismatches, are almost indistinguishable from those by a binary black hole. The disruption of the star significantly suppresses the kicks on the final black hole. The disruption also affects, although not dramatically, the spin of the final black hole. Lastly, for neutron stars with low compactness, the quasi-normal ringing of the black hole after the merger does not show a clean quasi-normal ringing because of the late accretion of debris from the neutron star. 
\end{abstract}

\section{Introduction}
The most recent catalogue of \gw{} observations (GWTC-3) from the LIGO-VIRGO-KAGRA (LVK) collaboration contain three possible \nsbh{} binary mergers (GW191219\_163120, GW200115\_042309, and GW200210\_092254)~\cite{PhysRevX.13.041039}. As the detectors increase their sensitivity, the number of \nsbh{}, and also \bns{} merger detections will increase. As with mergers of \bbh{s}, waveform from numerical simulations are important in characterizing the sources (masses, spins, eccentricity, etc.), and for the case of binaries with \ns{} components, simulations provide extremely valuable information about the internal structure of the \ns{} and the effects from tidal deformations and disruptions.

Several numerical studies of \nsbh{} mergers that include the effects of tidal deformations and disruption effects have been done in the past decade. 
To study the effect of mass ratio $q=M_h/M_*$ and compactness $C=M_*/R_*$ where $M_*$ and $R_*$ are respectively the mass and radius of the \ns{} and $M_h$ the mass of the \bh{}, Shibata \etal~\cite{shibata-2009} probed mass ratios $1.5 \le q \le 5$ and compactnesses $0.145 \le C \le 0.178$ and found that low compactness and mass ratio result in larger \ns{} disruption.
Duez \etal~\cite{Duez_2009} investigated a polytropic \eos{} $P=K \rho_0^\Gamma$ with adiabatic index $\Gamma=2,\ 2.75$ and two cases of nuclear-theory based Shen \eos{.} With fixed compactness of $C=0.15$, the cases varying \eos{} stiffness showed a limited effect on the total mass remnant, approximately $7\%$ of tidal debris outside \bh{} at the end of the simulation. However, the effect of stiffer \eos{} manifests in a more massive, larger, and longer-lived tidal tail. Kyutoku \etal~\cite{Kyutoku:2010zd} studied mixed binaries with non-spinning \bh{s} and piecewise-polytropic \eos{s}, with one piece for the core and another for the crust as given in~\cite{Read_2009}. Spinning \bh{s} were considered in~\cite{Kyutoku:2011vz}. The studies found the effects of the \eos{s} on tidal disruption in two aspects: First, \ns{s} with stiffer \eos{} have a larger radius and thus less compactness, yielding stronger disruption. Second, with fixed compactness, the comparison of different adiabatic index of the core showed that more centrally condensed density profile (higher adiabatic index) would result in a smaller mass remnant.
Etienne \etal~\cite{Etienne:2009} studied cases with aligned, anti-aligned, and zero \bh{} spins. The study found that massive disk ($\sim 0.2 M_{\odot}$) can be formed for the aligned and highly-spinning (dimensionless \bh{} spin $a=0.75$) case, whereas only small disks are formed for the anti-aligned cases. The effect of spin was investigated also by Kyutoku \etal~\cite{Kyutoku:2011vz}. The study found that for aligned cases, higher spin magnitudes enhances the \ns{} disruption and leads to more remnant mass. An enormous disk up to $0.5 M_{\odot}$ is obtained by Foucart \etal~\cite{Foucart_2011} as they pushed the \bh{} spin up to $a=0.9$. In the same study they also found that the misalignment angle has a strong affect on the remnant mass only for large angles $>40^{\circ}$ relative to the orbital angular momentum.
The magnetic field is also a factor affecting the remnant mass. Etienne \etal~\cite{Etienne_2012} performed full GRMHD simulations with a poloidal magnetic field. The study showed that the only cases where the magnetic field has significant impact on the remnant mass are when the maximum initial seeded field reaches $\sim 10^{17} G$. In a follow up study~\cite{Etienne_2012_tilted}, a tilted poloidal magnetic field in the \ns{} was shown  to lead to magnetic rotation instability, producing an outflow powerful enough to generate sGRBs. Using 26 numerical simulations, Foucart~\cite{Foucart:2012nc} constructed a model to predict the mass remnant by comparing the tidal radius to the ISCO radius of the system given its parameters.

In our previous study~\cite{Khamesra_2021}, we investigated the impact that the mass ratio has on the merger dynamics and the \gw{s} emitted for polytropic \eos{.} Specifically, we studied how as the mass ratio increases, the less mass ejecta is released, and the system behaves more like a \bbh{} system. As observed in Ref.~\cite{Shibata:2011jka}, the final fate of \nsbh{} binaries is determined by not only the mass ratio but also the compactness of the \ns{}. The main objective of the present work is to explore the effects of the compactness of the \ns{} from situations in which the \ns{} is deformed as it merges with the \bh{} to cases when the \ns{} is completely disrupted by the \bh{} before the merger. One motivation for carrying out this study is to test the ability of our initial data framework, an extension of the Bowen-York methods for \bbh{} merger, to model \nsbh{} systems. 

\begin{figure}[!htbp]
  \begin{center}
	\includegraphics[width=0.48\textwidth]{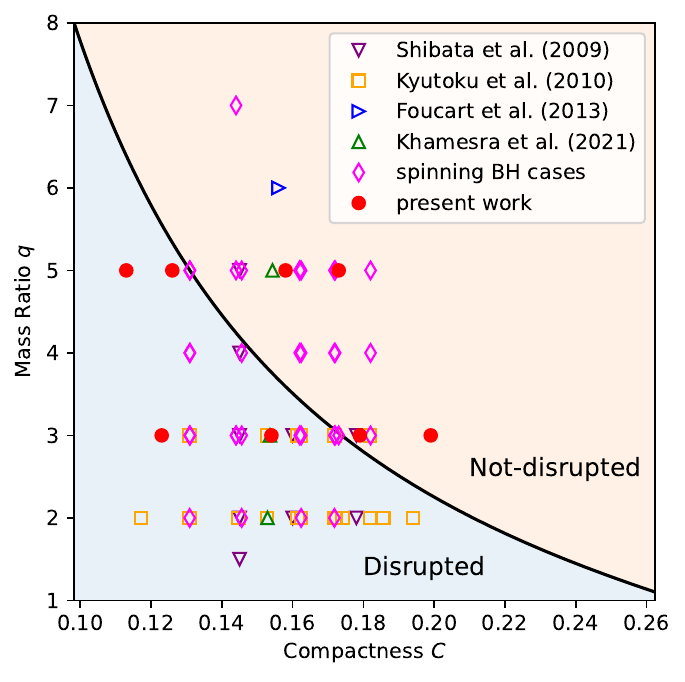}
	\caption[Critical Mass Ratio-Compactness Curve ]{The solid line denotes the boundary between \nsbh{} mergers in which the star is swallowed by the \bh{} almost undisturbed (orange region) and those binaries in which the star is disrupted before the merger (blue region).}
       \label{fig:ch5_QCcurve}
      \end{center}
\end{figure}

The parameter space under consideration is two-dimensional and consists of the mass ratio $q$ and compactness $C$. As depicted in Figure~\ref{fig:ch5_QCcurve}, there are two regions: values of $q$ and $C$ for which the star disrupts before the merger (blue region) and values for which the \ns{} remains basically intact before it is swallowed by the \bh{} (orange region). To get a rough estimate of the boundary separating these regions, we recall that the tidal redius $r_t$ is the separation at which the tidal force $M_h\,M_*\,R_*/r_t^3$ by the \bh{}  equals the star's self-gravity $M_*/R_*^2$. This yields $C^3=q^{-2}\,(M_h/r_t)^3$. A good approximation to the minimum tidal radius for which the \ns{} disrupts is the radius of the inner-most stable circular orbit (ISCO), which for a point particle orbiting a non-rotating \bh{} is $6\,M_h$. Thus, the boundary between disruption and non-disruptions is given by
$(6\,C)^{3/2}\,q = 1$. A more accurate expression, which is depicted in Fig.~\ref{fig:ch5_QCcurve} with a black line,  was obtained by Taniguchi \etal \cite{Taniguchi:2007aq} using results from numerical simulations. The expression reads
\begin{equation}
    (6\, C)^{3/2} \,q = 3.7\,(1+1/q)^{-3/2}\left[ 1-0.444\,q^{-1/4}\left(1-3.54\, C^{1/3}\right)\right]\,.
    \label{eq:rtidal}
\end{equation}

In Fig.~\ref{fig:ch5_QCcurve}, with green triangles we denote the simulations in our previous work~\cite{Khamesra_2021}, and with red dots the simulations for the present study. The figure also includes the simulations by Kyutoko \etal ~\cite{Kyutoku:2010zd}, Shibata \etal \cite{shibata-2009}, Foucart \etal \cite{Foucart:2013psa}, and cases with spinning black holes \cite{Etienne:2009,Kyutoku:2011vz,Foucart_2011,Foucart_2013}. 
That is, our present study considers the regime of $0.11$ to $0.20$ compactness for mass ratio $q=3$ and $q=5$ to find distinguishing properties separating disrupted and non-disrupted mixed binaries.

The paper is organized as follows: A summary of the method to construct initial data for binary systems with \ns{s} is presented in Section~\ref{sec:init_data}. Numerical setup, simulation parameters, and convergence tests are given in Section~\ref{sec:initparams_nrsetup}. Results are presented and discussed in Section~\ref{sec:results}, with conclusions in Section~\ref{sec:conclusions}. Quantities are reported in units of $M=M_h+M_*$, with $G = c = 1$. Space-time signature is $(-+++)$, and tensor indices are denoted with Latin letters from the beginning of the alphabet. Spatial tensor indices are denoted with Latin letters from the middle of the alphabet.

\section{Initial Data for Binary Systems with Neutron Stars}
\label{sec:init_data}
In Ref.~\cite{Clark:2016ppe}, we introduced a method to construct initial data for \nsbh{} binaries following parallel steps to the Bowen-York approach for \bbh{} initial data with \bh{s} modeled as punctures. The initial data is comprised of the spatial metric $\gamma_{ij}$ and the extrinsic curvature $K_{ij}$ of the initial space-like hypersurface. For the matter sources, the data involves 
\begin{eqnarray}
\label{eq:hmdensity_pf}
\rho_H &\equiv n^a n^b T_{ab}\\
\label{eq:momdensity_pf}
S^i &\equiv -\gamma^{ij} n^b T_{jb}\,,
\end{eqnarray}
where $T_{ab}$ is the stress-energy tensor and  $n^a$ is the unit time-like normal to the hypersurface. $\rho_H$ and $S^i$ are respectively the energy and momentum density measured by normal $n^a$ observers. We consider a perfect fluid for which 
\begin{equation}
\label{eq:stressenergy}
T_{ab} = (\rho + p)u_a u_b + p\, g_{ab},
\end{equation}
with $\rho$ the total energy density, $p$ the pressure, $u^a$ the four velocity of the fluid, and $g_{ab} = \gamma_{ab}-n_an_b$ the space-time metric. With this, 
\begin{eqnarray}
\label{eq:hmdensity_pf2}
\rho_H &=&  (\rho + p)\,W^2 - p\\
\label{eq:momdensity_pf2}
S^i &=&  (\rho + p) \,W\,u^i,
\end{eqnarray}
where $W = -n_au^a$ is the Lorentz factor, also given by
\begin{equation}
        W^2 = \frac{1}{2}\left(1+\sqrt{1 + \frac{4 S_iS^i}{(\rho + p)^2}}\right).
\end{equation}
The initial data $\lbrace \gamma_{ij}, K_{ij}, \rho_H, S^i \rbrace$ is not freely specifieable. The data must satisfy
\begin{eqnarray}
\label{eq:ham}
    R + K^2 - K_{ij} K^{ij} &= 16 \pi \rho_H\\
\label{eq:mom}
    \nabla_j \left(K^{ij} - \gamma^{ij}K \right) &= 8 \pi S^i,
\end{eqnarray}
which are respectively the Hamiltonian and Momentum constraints. In these equations, $R$ is the Ricci scalar and $\nabla_j$ is the covariant derivative associated with $\gamma_{ij}$.

Eqs.(\ref{eq:ham}) and (\ref{eq:mom}) are solved using the conformal-transverse-traceless (CTT) approach pioneered by Lichnerowicz~\cite{baumgarte_shapiro_2010}, York, and collaborators~\cite{Smarr:1979ofa} in which 
\begin{eqnarray}
    \gamma_{ij} &=& \psi^4 \tilde{\gamma}_{ij}\\
    K_{ij} &=& A_{ij} + \frac{1}{3}\gamma_{ij}K \\
    A^{ij} &=&\psi^{-10} \tilde{A}^{ij}\\
    \tilde{\rho}_H &=& \rho_H \psi^8\\
     \tilde{S}^i &=& S^i \psi^{10}\,.
\end{eqnarray}
From the last two transformations, we get that $\tilde{\rho} = \rho\, \psi^8$, $\tilde{p} = p\, \psi^8$, $\tilde{u}^i = u^i\psi^2$ and $\widetilde W = W$.

As commonly done, we impose conformal flatness ($\tilde{\gamma}_{ij} = \eta_{ij}$), maximal slicing ($K=0$), and $\tilde{A}^{\rm TT}_{ij} = 0$. Thus, the Hamiltonian and momentum constraints become:
\begin{eqnarray}
    \label{eq:ham2}
    &\tilde{\Delta} \psi+ \frac{1}{8}\psi^{-7}\tilde{A}_{ij}\tilde{A}^{ij} = -2\pi \psi^{-3}\tilde{\rho}_H \\
    \label{eq:mom2}
    &\widetilde{\nabla}_j \tilde A^{ij} = 8 \pi \tilde{S}^i.
\end{eqnarray}
Bowen and York ~\cite{1980PhRvD..21.2047B} found that a point-source (puncture) solution to $\widetilde{\nabla}_j \tilde A^{ij} = 0$ representing \bh{s} with linear momentum $P^i$ is given by:
\begin{eqnarray}
  \tilde A^{ij} &=& \frac{3}{2\,r^2}\left[ 2\,P^{(i} l^{j)} - ( \eta^{ij}-l^il^j)P_k l^k\right]\label{eq:KP}
\end{eqnarray}
where $l^i = x^i/r$ is a unit radial vector. Bowen~\cite{Bowen1979} also found the following solution to Eq.~(\ref{eq:mom2}) for a spherically symmetric source $S^i$ representing an extended object with linear momentum $P^i$:
\begin{eqnarray}
\tilde{A}^{ij} &=& \frac{3Q}{2r^2}\left[2P^{(i}l^{j)} - (\eta^{ij} - l^il^j)P_kl^k  \right] \nonumber\\
&+& \frac{3C}{r^4}\left[2P^{(i}l^{j)} + (\eta^{ij} - 5l^il^j)P_kl^k  \right]\label{eq:KP2}
\end{eqnarray}
where
\begin{eqnarray}
	\label{eq:qjcdefs}
	Q &=&4\pi \int_0^r  \sigma \bar r^2 \, d\bar r \\
	C &=& \frac{2\pi}{3} \int_0^r  \sigma \bar r^4 \, d\bar r\,.
\end{eqnarray}
The source function $\sigma$ is a radial functions with compact support $r\le R$ such that
$\tilde{S}^i = P^i \sigma$.
Outside the source, $Q=1$ and $C=0$, and the extrinsic curvature (\ref{eq:KP2}) reduces to  (\ref{eq:KP}). Since $\tilde S^i = (\tilde\rho+\tilde p) W\,\tilde u^i $, we set
$\sigma  = (\tilde\rho+\tilde p)/\mathcal K$ with 
\begin{eqnarray}
   {\mathcal K} &=&  4\,\pi \int_{0}^{R} (\tilde\rho+\tilde p)\,r^2\, dr\,.
\end{eqnarray}
Thus, $P^i = W\,{\mathcal K}\,\tilde u^i$.

Given (\ref{eq:KP2}) for a \ns{} and (\ref{eq:KP}) for a \bh{,} we solve the Hamiltonian constraint (\ref{eq:ham2}), assuming that the conformal factor has the form $\psi = 1+m_p/(2r)+u$ where $m_p$ is the bare or puncture mass of the \bh{.} To solve (\ref{eq:ham2}), we use  the \texttt{TwoPunctures} code~\cite{Ansorg:2004ds}, that has been modified to handle the source $\tilde\rho_H$.
To obtain initial data representing a \nsbh{} binary system with a \bh{} with irreducible mass $M_h$ and a \ns{} with mass $M_*$, one follows similar steps to that for \bbh{s}.
That is, one selects the target values for $M_*$ and $M_h$. Next, one carries out  iterations solving the Hamiltonian constraint until the target values for $M_h$ and $M_*$ are obtained. After each Hamiltonian constraint solve iteration, one computes $M_h$ from the irreducible mass of the \bh{} from its horizon. The challenge is in finding an appropriate definition for $M_*$. Options are the ADM mass $\mathcal{M}_A$ or the rest mass $\mathcal{M}_0$  of the \ns{} at rest in isolation, which in isotropic coordinates read
\begin{eqnarray}
	\mathcal M_{A} &=& 4\,\pi \int_0^R\rho\, \psi^5 r^2\, dr \\
    \mathcal M_0 &=&  4\,\pi \int_0^R  \rho_0\, \psi^{6} r^2\, dr
\end{eqnarray}
respectively, with $\rho_0$ the rest-mass density.
The approach followed in Ref.~\cite{Clark:2016ppe} was to compute $M_*$ at iteration $n$ from $M_*^{(n)} = \xi^{(n-1)}\,M_0^{(n)}$ where $\xi^{(n-1)} = \mathcal M_A^{(n-1)}/\mathcal M_0^{(n-1)}$ and
\begin{equation}
    M_0 = \int  \rho_0 W \sqrt{\gamma} d^3x =  \int  \tilde\rho_0 W \psi^{-2} d^3x\,.
\end{equation} 
At $n=1$, we set $M_*^{(1)} =\mathcal M_A^{(1)}$. At the end of the iterations, we obtain the rest mass density and puncture mass that generates desired $M_h$ and $M_*$.

\section{Simulations parameters, numerical setup, and convergence tests}
\label{sec:initparams_nrsetup}

 We set the target mass of the \ns{} in the binary to $M_* = 1.35\,M_\odot$. With this mass and the range of compactness $0.113 \le C \le 0.199$, the \ns{s} have radius $10.2\,\textnormal{km} \le R_* \le 17.8\,\textnormal{km}$. Also, with this \ns{} mass, the \bh{} has a mass $M_h = 4.05\,M_\odot$ for $q=3$ and $M_h = 6.75\,M_\odot$ for $q=5$. We model the \ns{} as a polytrope with an equation of state $P=K \rho_0^\Gamma$ with $\Gamma = 2$. 
 Table \ref{tab:table_id_config_ch5} provides the \ns{} radius $R_*$, central density $\rho_c$, ADM mass $\mathcal{M}_A$ and rest mass $\mathcal{M}_0$ of the \ns{} in isolation, as well as the polytropic constant $K$ and the grid resolution $\Delta$ inside the \ns{} for each compactness case.
At the beginning of the simulation, all the binary systems we consider have a coordinate separation of $9\,M$. As with \bbh{} initial data, the initial momentum of the \bh{} and \ns{} are obtained by integrating the post-Newtonian equations of motion from a large separation and stopping the integration when the binary separation is $9\,M$.  

\begin{table}[!htbp]
\centering
 \begin{tabular}{c c c c c c c}
\toprule
	 \tabhead{$C$} & \tabhead{$R_*(km)$} & \tabhead{$\rho_c(gr/cm^3)$} & \tabhead{$\mathcal{M}_A (M_\odot)$} & \tabhead{$\mathcal{M}_{0}(M_\odot)$} & \tabhead{$K(km^2)$} & \tabhead{$\Delta (m)$}\\
 \midrule
          \midrule
          &        \multicolumn{5}{c}{$q =3$} \\ 
          \midrule
        0.123 & 16.834 & 4.65e+14 & 1.398 & 1.487 & 276.101 & 302.53 \\
        0.152 & 13.492 & 9.11e+14 & 1.393 & 1.503 & 204.796 & 236.52 \\
        0.179 & 11.459 & 1.51e+15 & 1.389 & 1.514 & 172.085 & 192.52 \\
        0.199 & 10.285 & 2.13e+15 & 1.385 & 1.519 & 159.252 & 167.77 \\
             \midrule
          &        \multicolumn{5}{c}{$q =5$} \\ 
          \midrule
        0.113 & 18.728 & 3.44e+14 & 1.428 & 1.512 & 327.231 & 346.54 \\
        0.126 & 16.699 & 4.87e+14 & 1.427 & 1.521 & 276.101 & 305.28 \\
        0.158 & 13.263 & 9.83e+14 & 1.423 & 1.539 & 204.306 & 235.15 \\
        0.173 & 12.114 & 1.30e+15 & 1.420 & 1.545 & 185.431 & 210.40 \\
\bottomrule \\
\end{tabular}
\caption{Compacness $C$, \ns{} radius $R_*$, central density $\rho_c$, ADM mass $\mathcal{M}_A$ and rest mass $\mathcal{M}_0$ of the \ns{} in isolation, as well as the polytropic constant $K$ and the grid resolution $\Delta$ inside the \ns{}.}
\label{tab:table_id_config_ch5}
\end{table}

The system of spacetime and hydrodynamics equations are solved with \texttt{MAYA} code~\cite{2015ApJLEvans,2016PRDClark,2016CQGJani}, our local version of the \texttt{Einstein Toolkit} code~\cite{EinsteinToolkit:2023_11}. The Einstein equations are solved with the BSSN-Chi formulation~\cite{Shapiro1999,Shibata1995,Beyer:2004sv}. The general relativistic hydrodynamics equations are solved following the formulation in the \texttt{Whisky} public code~\cite{Baiotti:2004wn,Hawke:2005zw, Baiotti:2010zf}. To evolve the spacetime, we use the moving puncture gauge~\cite{Campanelli2005,Baker2006}.
The \bh{} is tracked with the \texttt{AHFinderDirect} code~\cite{Thornburg:2003sf}. The \ns{} is tracked via the \texttt{VolumeIntegrals} thorn~\cite{EinsteinToolkit:2023_11}, which locates the \ns{} from the center of mass within a box encasing the star. The mass, spins, and multipole moments of the \bh{} are computed using the \texttt{QuasiLocalMeasures} thorn~\cite{Loffler:2011ay} based on the dynamical horizons framework~\cite{2004LRR.....7...10A}. The \gw{} strain is obtained from the Weyl scalar $\Psi_4$~\cite{Loffler:2011ay, Zilhao:2013hia,Reisswig:2010di}. To compute the radiated quantities, we follow the method developed in \cite{Ruiz:2007yx}. The calculations of $\Psi_4$, the strain, and the radiated quantities are computed using the Python library \texttt{mayawaves}~\cite{ferguson_2023mayawaves}.

We use the moving box mesh refinement approach as implemented by \texttt{Carpet}~\cite{2016ascl.soft11016S}. To ensure uniformity of the resolution across the runs, we follow conditions based on the analysis in Ref.~\cite{Shibata:2011jka}, which suggested that the grid spacing inside the \bh{} should be smaller than $\Delta \sim M_h/20$ and inside the \ns{} smaller than $\Delta \sim R_*/40$. To meet this requirement, we use $8$ levels of refinement on the star. The star is covered by $96^3$ points, with $116^3$ points in the coarsest level. The finest refinement has an extent of $2.4\, R_*$. At the \bh{} we add another level of refinement. For wave extraction, we set up a sphere at radius $130\,M$ from the center of mass of the binary to capture $\Psi_4$ and the radiated quantities.

 \begin{figure}[!htbp]
  \begin{center}
 	\includegraphics[width=0.95\textwidth]{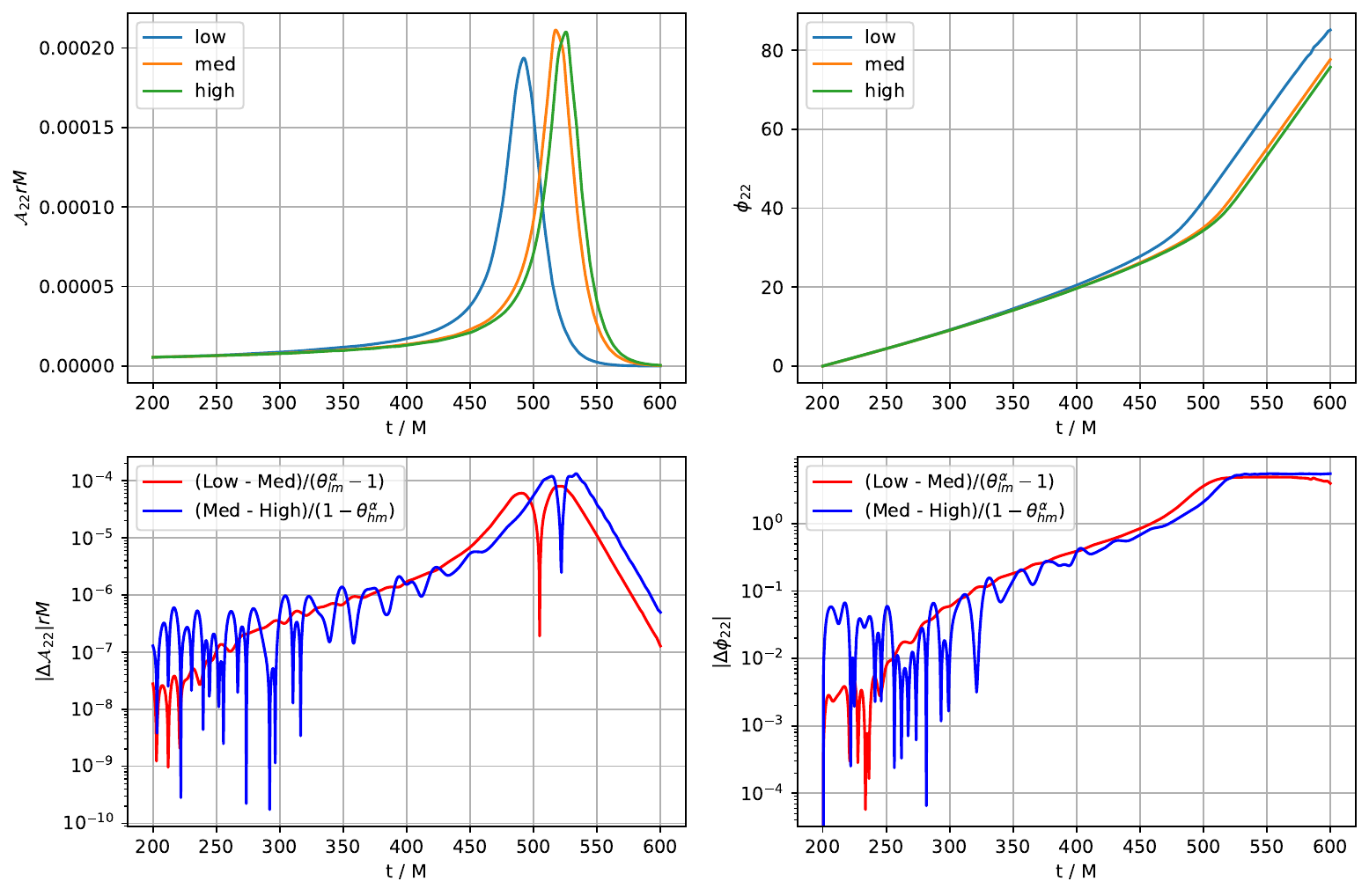}
 	\caption[Convergence of mass ratio $q=5$ mixed binary.]{Convergence results for the (2,2) mode of the Weyl Scalar $\Psi_4$ for mass ratio $q=5$ and compactness $C=0.173$. The top panels show the amplitude (left) and phase evolution (right) for three simulations with resolutions at the finest level of $\Delta_{h}=188$ meters (high), $\Delta_{m}=210$ meters (medium), and $\Delta_{l}= 339$ meters (low). The bottom panels show the left and right-hand side of equation (low - medium)$/(\theta_{lm}^\alpha - 1)$ = (high - medium)$/(1- \theta_{hm}^\alpha)$ with  $\theta_{lm} = \Delta_{l}/\Delta_{m}$ and $\theta_{hm} = \Delta_{h}/\Delta_{m}$. In the bottom left for the amplitude we used $\alpha=4.0$ and in the bottom right panel for the phase $\alpha=3.97$.}
        \label{fig:ch3_convergence}
       \end{center}
 \end{figure}

For convergence tests, we focus on the most compact case $C=0.173$ for mass ratio $q=5$. Our previous work~\cite{Khamesra_2021} showed convergence tests for $q=2$ and $3$.
In this work, we carried out three simulations with resolutions in the finest mesh at the \ns{} of $\Delta_{h}=188$ meters (high), $\Delta_{m}=210$ meters (medium), and $\Delta_{l}= 339$ meters (low). Top panels in Figure \ref{fig:ch3_convergence} show the amplitude (left) and phase (right) for the three simulations. Assuming a convergence rate of $\alpha$, one should have that (low - medium)$/(\theta_{lm}^\alpha - 1)$ = (high-medium)$/(1- \theta_{hm}^\alpha)$ with $\theta_{lm} = \Delta_{l}/\Delta_{m}$ and $\theta_{hm} = \Delta_{h}/\Delta_{m}$. The bottom panels show the left and right-hand sides of this equation for the amplitude (left panel) with $\alpha=4.0$ and the phase (right panel) with $\alpha=3.97$. This is consistent with the 4-th order convergence expected from the order of discretization and temporal updating in the code.

\section{Results}
\label{sec:results}

\begin{figure}[!htbp]
\centering
    \begin{subfigure}{0.49\textwidth}
        \centering
        \includegraphics[width=0.98\linewidth]{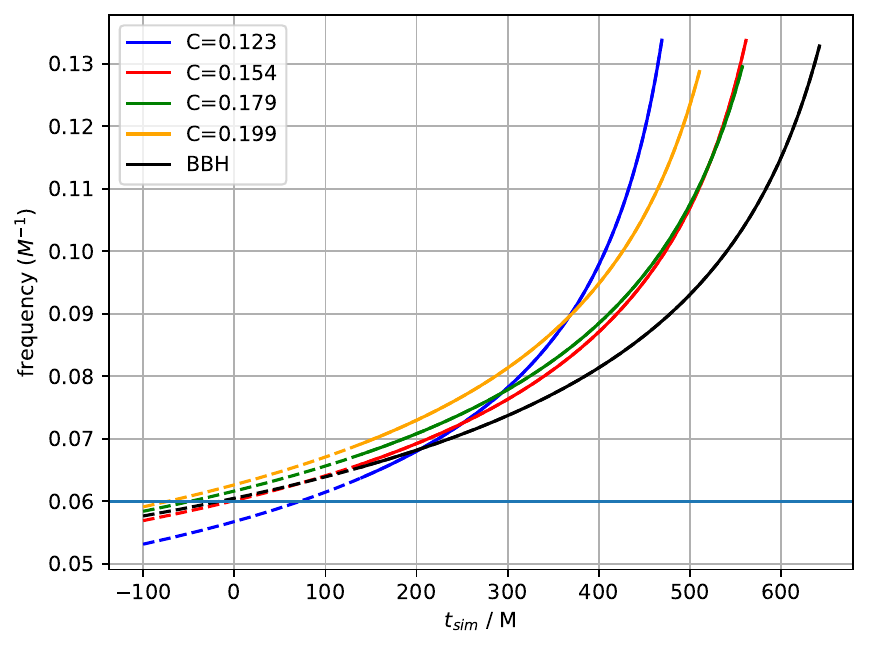}
        \caption{Mass ratio $q=3$.}
        
    \end{subfigure}
    \begin{subfigure}{0.49\textwidth}
        \centering
        \includegraphics[width=0.98\linewidth]{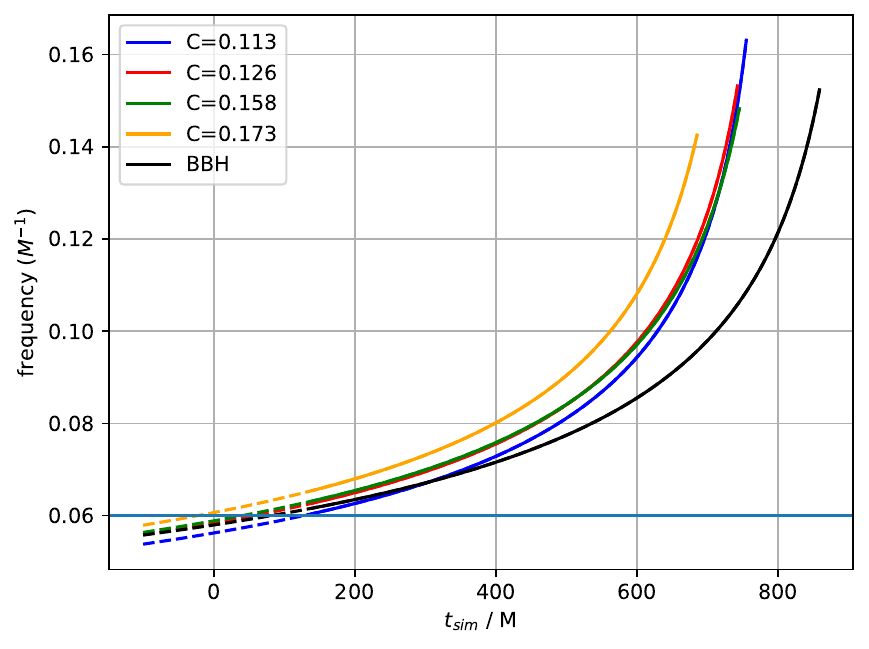}
        \caption{Mass ratio $q=5$.}
    \end{subfigure}
    \caption[The fitted frequency of gravitational waves.]{Frequency of the \gw{} as a function of simulation time. 
Solid lines are polynomial fits to the frequencies from the simulations after the junk radiation has passed, approximately at $140\,M$ from the beginning of the simulation. The lines end at $100\,M$ before the merger. The dashed lines are extrapolations back in time.}
\label{fig:ch5_freq_std}
\end{figure}

A direct comparison of waveforms from the start time of the simulations could introduce spurious nonphysical differences due to gauge effects. All the simulations in our study were done using the moving puncture gauge, but this does not necessarily imply that the time coordinate in the simulations are {\it aligned}. This is because of the differences in the parameters of the binaries. This can be seen in Figure~\ref{fig:ch5_freq_std} where we plot the frequency of the \gw{} as a function of simulation time. 
Solid lines are polynomial fits to the frequencies from the simulations after the junk radiation has passed~\cite{PhysRevD.100.081501}, approximately at $140\,M$ from the beginning of the simulation. The lines end at $100\,M$ before the merger. The dashed lines are extrapolations back in time. 
We follow the suggestion in Ref.~\cite{Foucart:2013psa} and shift the time coordinate in the simulation results such that $t = 0\,M$ is the time at which the \gw{} has a specified frequency. We picked this frequency to be $0.06\, M^{-1}$.

\subsection{Inspiral and Tidal Disruption}

\begin{figure}[!htbp]
\includegraphics[width=0.95\textwidth]{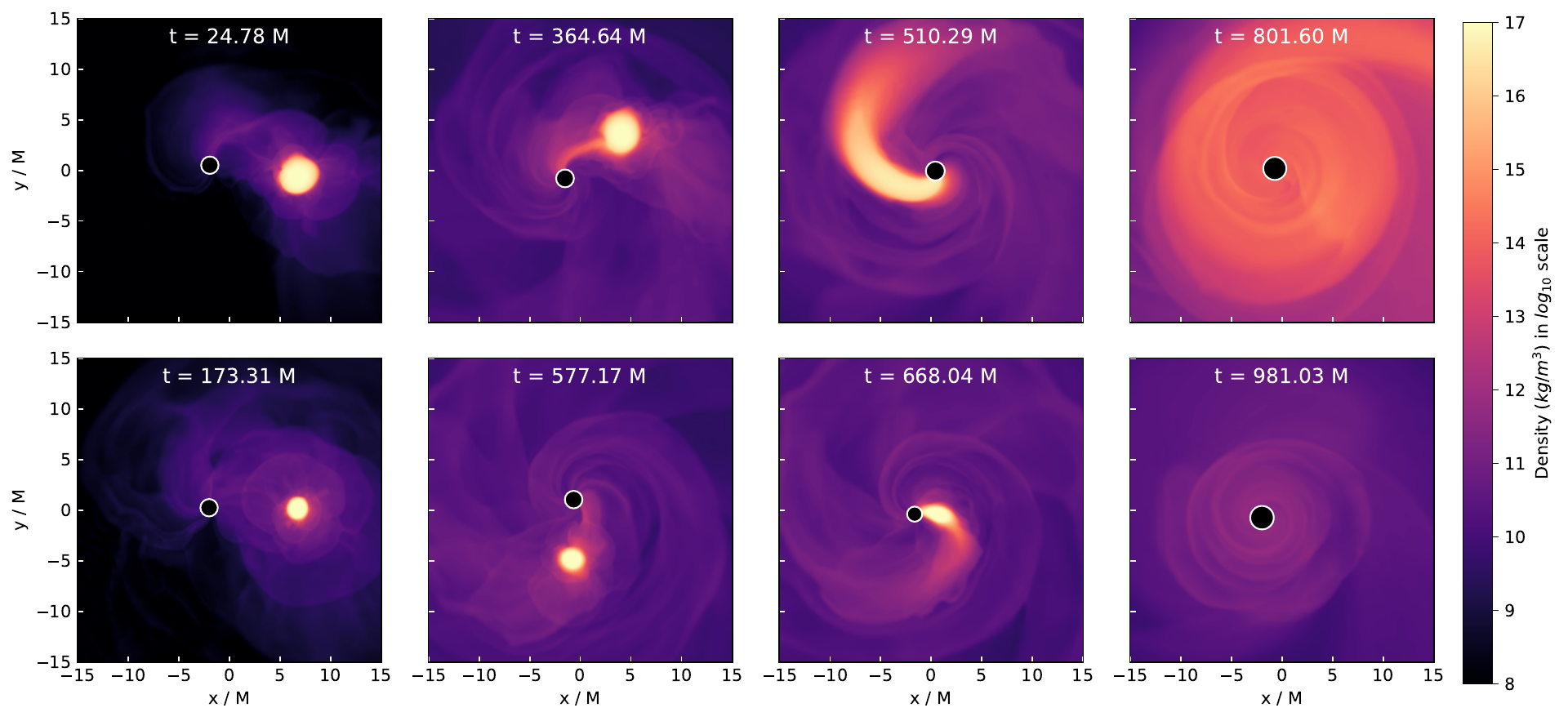}
\centering
	\caption{Rest mass density contour snapshots in the equatorial plane for two cases with mass ratio $q=3$. The top panel shows the case $C=0.123$, and the bottom panels are for $C=0.199$. The white circle denotes the black hole's apparent horizon.}
\label{fig:ch5_densitysnapshots_q3}
\end{figure}

\begin{figure}[!htbp]
\includegraphics[width=0.95\textwidth]{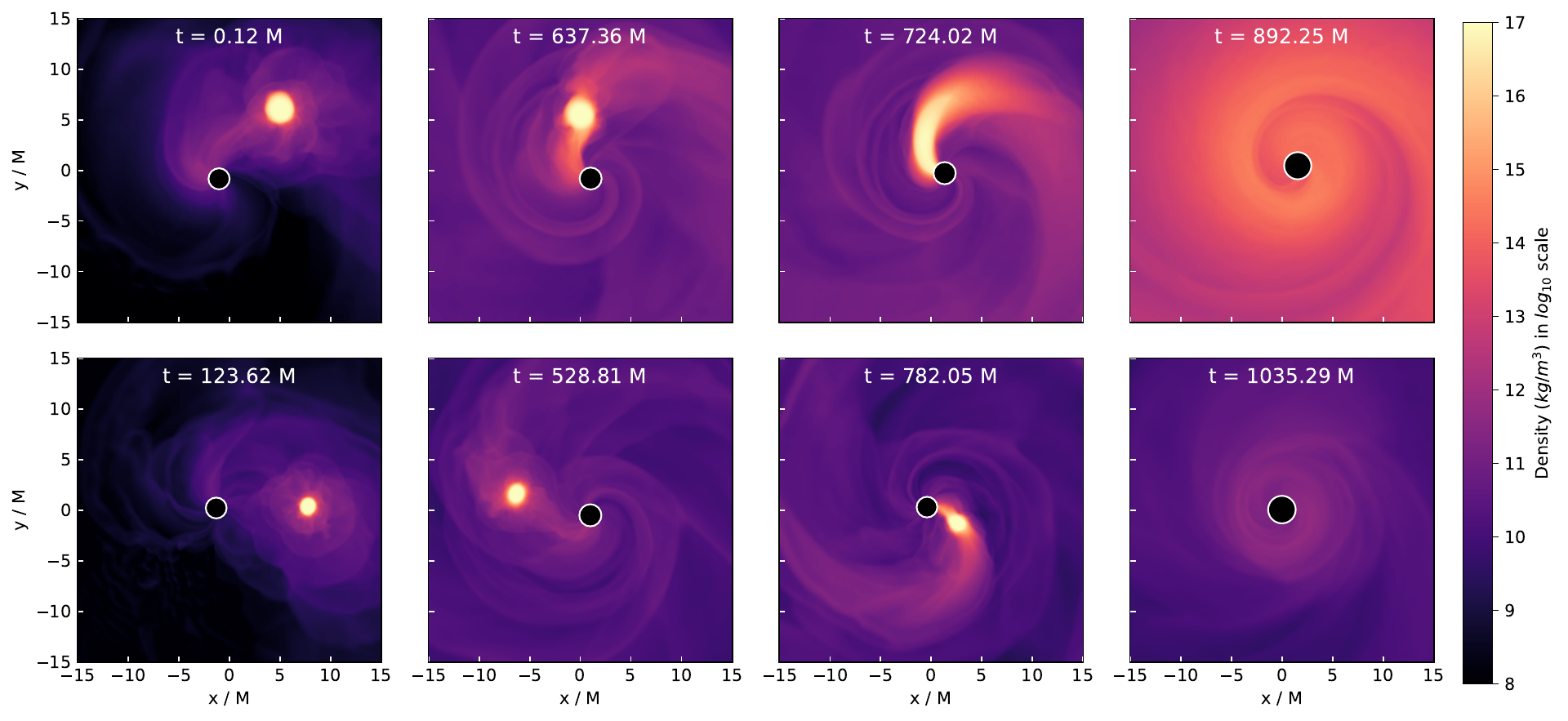}
\centering
	\caption{Rest mass density contour snapshots in the equatorial plane for two cases with mass ratio $q=5$. The top panel shows the case $C=0.113$, and the bottom panels are for $C=0.173$. The white circle denotes the black hole's apparent horizon.}
\label{fig:ch5_densitysnapshots_q5}
\end{figure}

Figures~\ref{fig:ch5_densitysnapshots_q3} and~\ref{fig:ch5_densitysnapshots_q5} show snapshots of the rest mass density in the equatorial plane for two cases with $q=3$ and another two for $q=5$, respectively. The top panels in Fig.~\ref{fig:ch5_densitysnapshots_q3} show the least compact case with $C=0.123$ and the bottom panels for the most compact case $C=0.199$. Similarly, the top panels in Fig.~\ref{fig:ch5_densitysnapshots_q5} show the case $C=0.113$ and the bottom panels with $C=0.173$. The white circle denotes the black hole's apparent horizon in all snapshots.

The first thing to notice in Figs.~\ref{fig:ch5_densitysnapshots_q3} and~\ref{fig:ch5_densitysnapshots_q5} is that, when comparing the panels for the high compactness cases (bottom panels) between $q=3$ and $q=5$, they show very similar qualitatively features. In both cases, the \ns{} is swallowed by the \bh{}, experiencing small disruption and mass loss. However, if one pays attention to the time stamp in the snapshots, one sees that the features in the $q=5$ case develop later. This is because the energy emitted in \gw{} scales roughly as $q^2/(1+q)^4$, and thus the luminosity of the $q=3$ case is larger than in $q=5$, i.e., the $q=5$ binary mergers later. 

The low compactness cases also show similar qualitative features and a time delay in $q=5$  compared to $q=3$. Here, there is an additional factor besides the one mentioned before depending on $q$. The tidal radius scales with compactness and mass ratio as 
$r_t \propto C^{-1}\,q^{-2/3}$. Thus, $r_t$ in the case $q=5$ is smaller than for $q=3$, and the star is able to inspiral longer before tidal forces from the hole disrupt the star. Nonetheless, before disrupting, the star remains fairly stable, losing less than $0.2 \%$ of the initial mass. As the star disrupts, it develops a spiral, fan-looking shape, a tail characteristic of tidal disruptions. The last two snapshots in the top panels of the figures show the late stage. In them, one can observe that as the tidal debris circularizes, there are hints of an accretion disk. The accretion in this stage will have, as we will discuss in a subsequent section, a profound effect on the quasi-normal ringing of the \bh{.} 

Finally, we observe that as the compactness increases, the mass shedding of the star and formation of the tidal tail occurs closer to the merger, as expected. This is consistent with the results from the study of quasi-equilibrium states of mixed binaries~\cite{Taniguchi:2007aq} in which it was found that frequency when shedding initiates is  $\propto C^{3/2}(1+1/q)^{1/2}$.

\subsection{Gravitational waves}
\label{sec:results-GW}
\begin{figure}[!htbp]
\centering
\includegraphics[width=\textwidth]{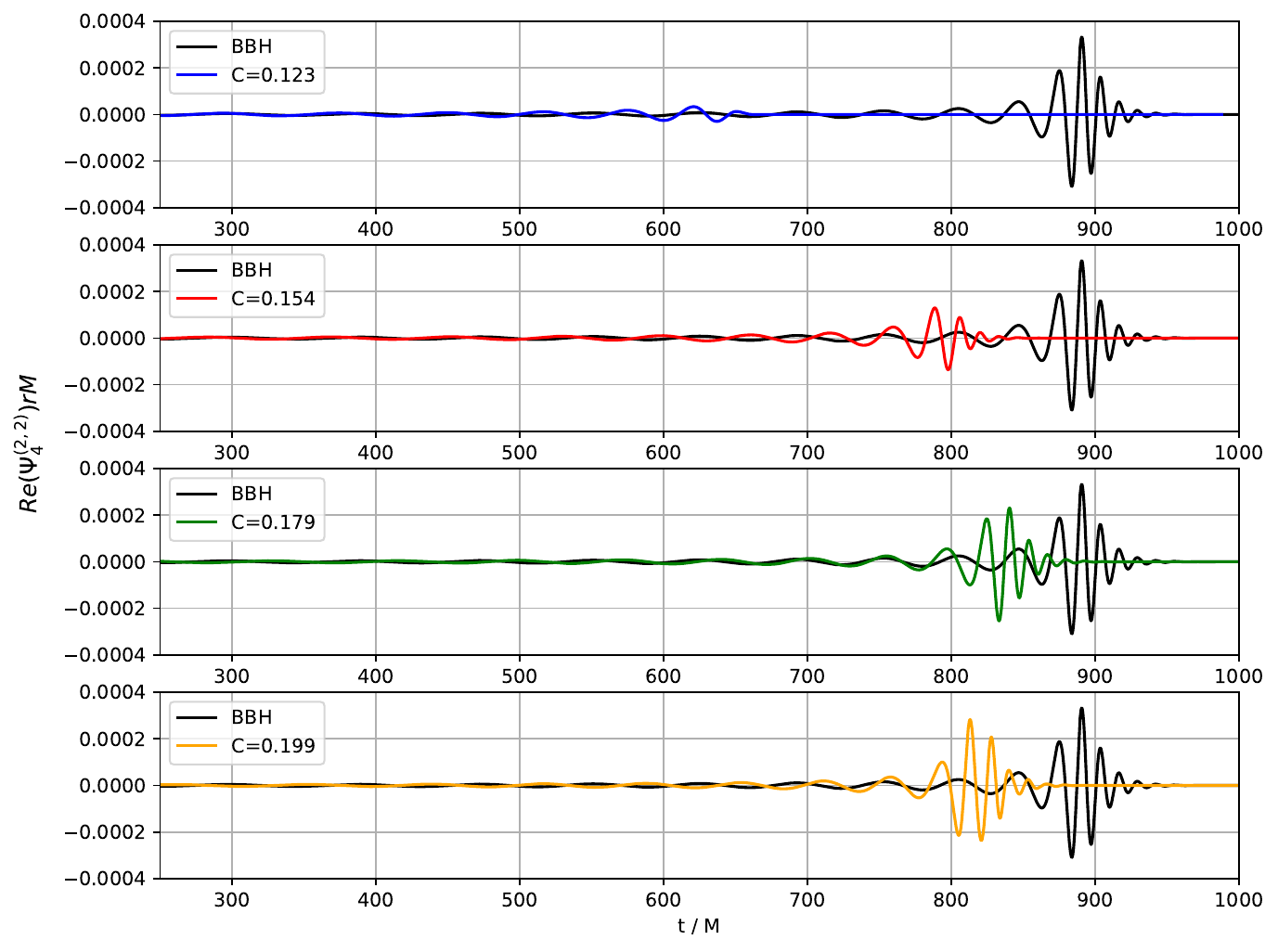}
	\caption{Comparison of $Re(\Psi_4^{(2,2)})$ for the $q=3$ case between \bbh{} and \nsbh{} simulations. Panels from top to bottom are in increasing compactness cases.}
\label{fig:psi4_22_q3}
\end{figure}

\begin{figure}[!htbp]
\centering
\includegraphics[width=\textwidth]{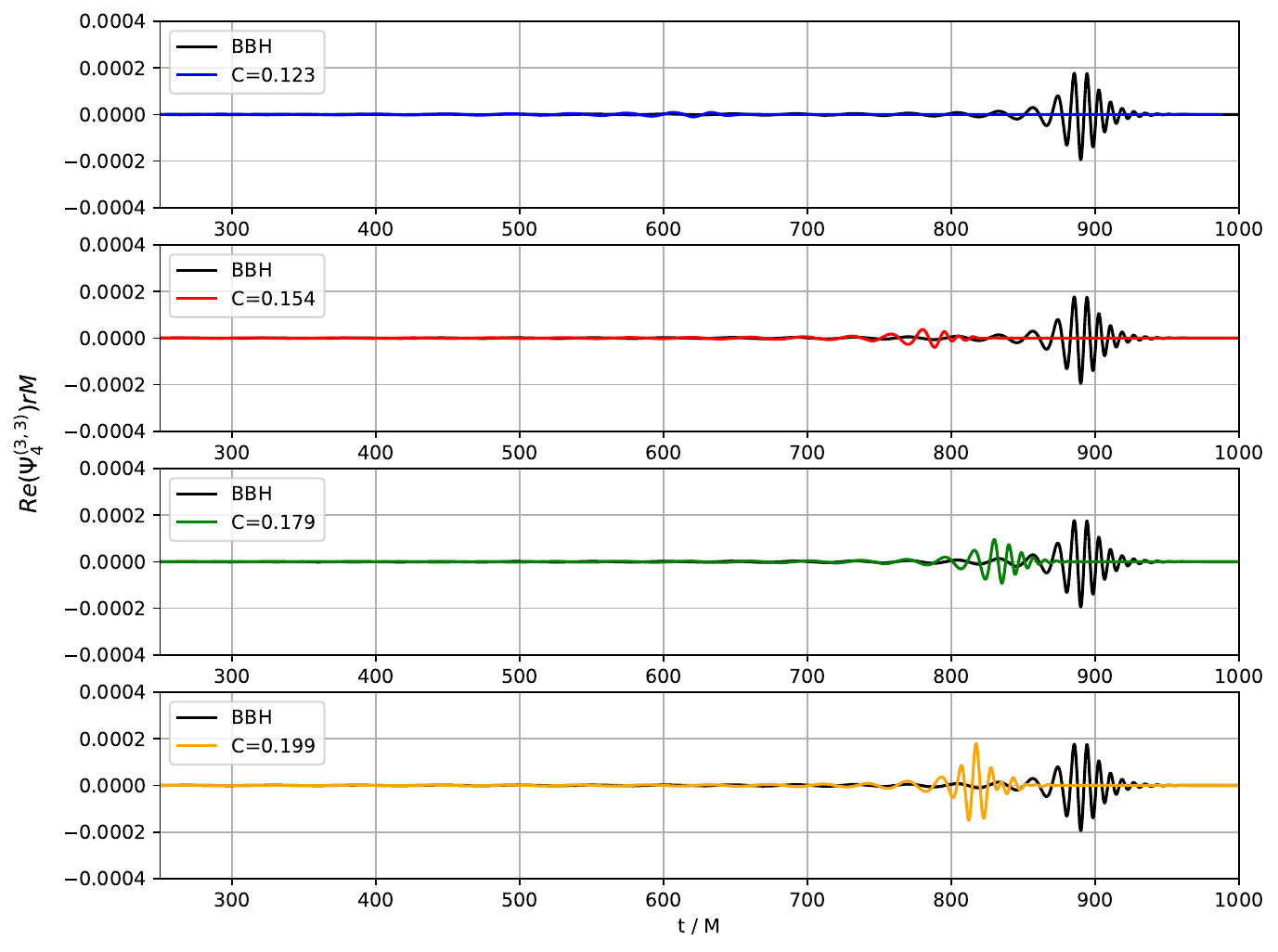}
	\caption{Same as in Fig.~\ref{fig:psi4_22_q3} but for the mode $Re(\Psi_4^{(3,3)})$.}
\label{fig:psi4_33_q3}
\end{figure}

\begin{figure}[!htbp]
\centering
\includegraphics[width=\textwidth]{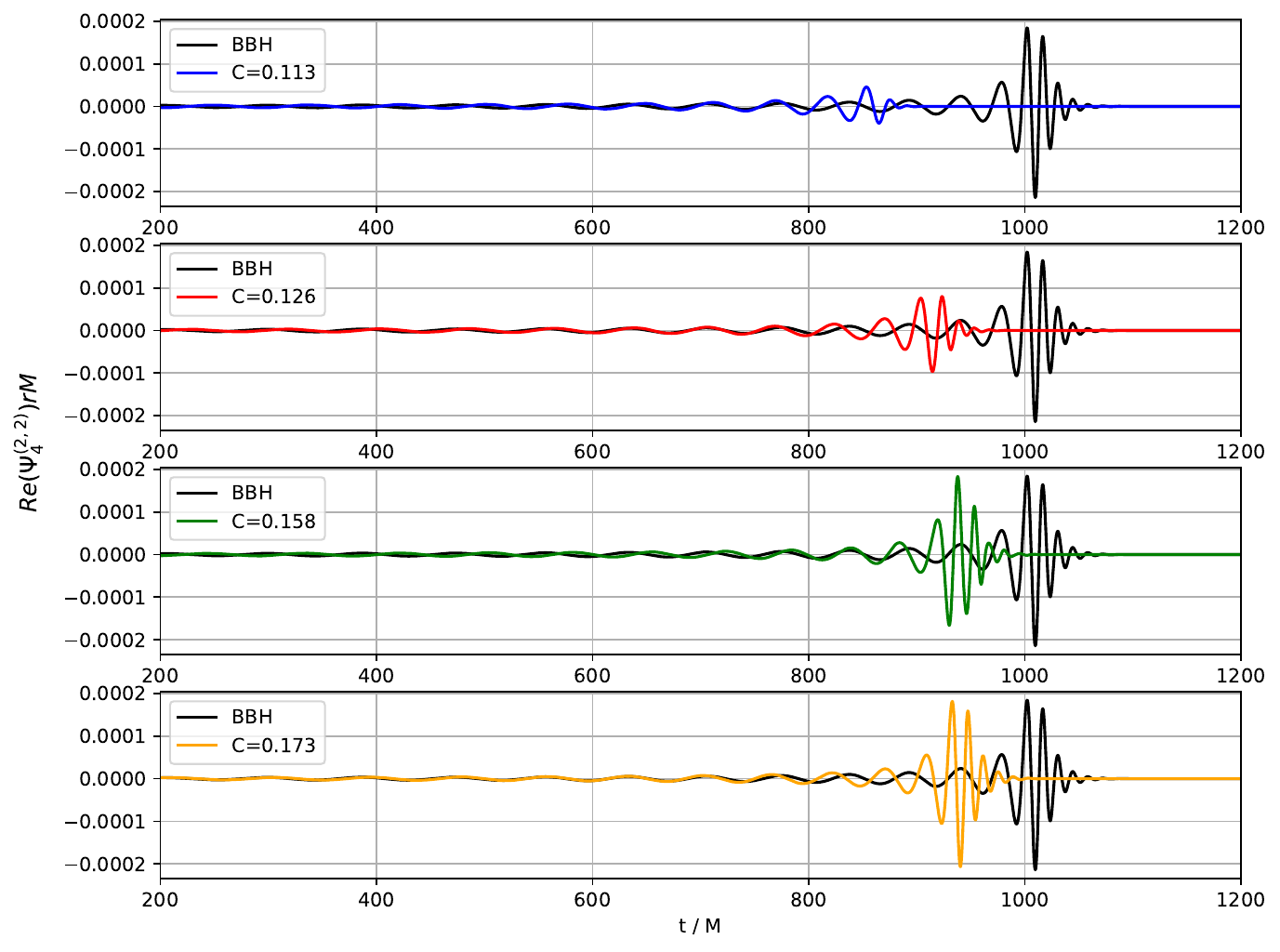}
	\caption{Same as in Fig.~\ref{fig:psi4_22_q3} but for the case $q=5$.}
\label{fig:psi4_22_q5}
\end{figure}

\begin{figure}[!htbp]
\centering
\includegraphics[width=\textwidth]{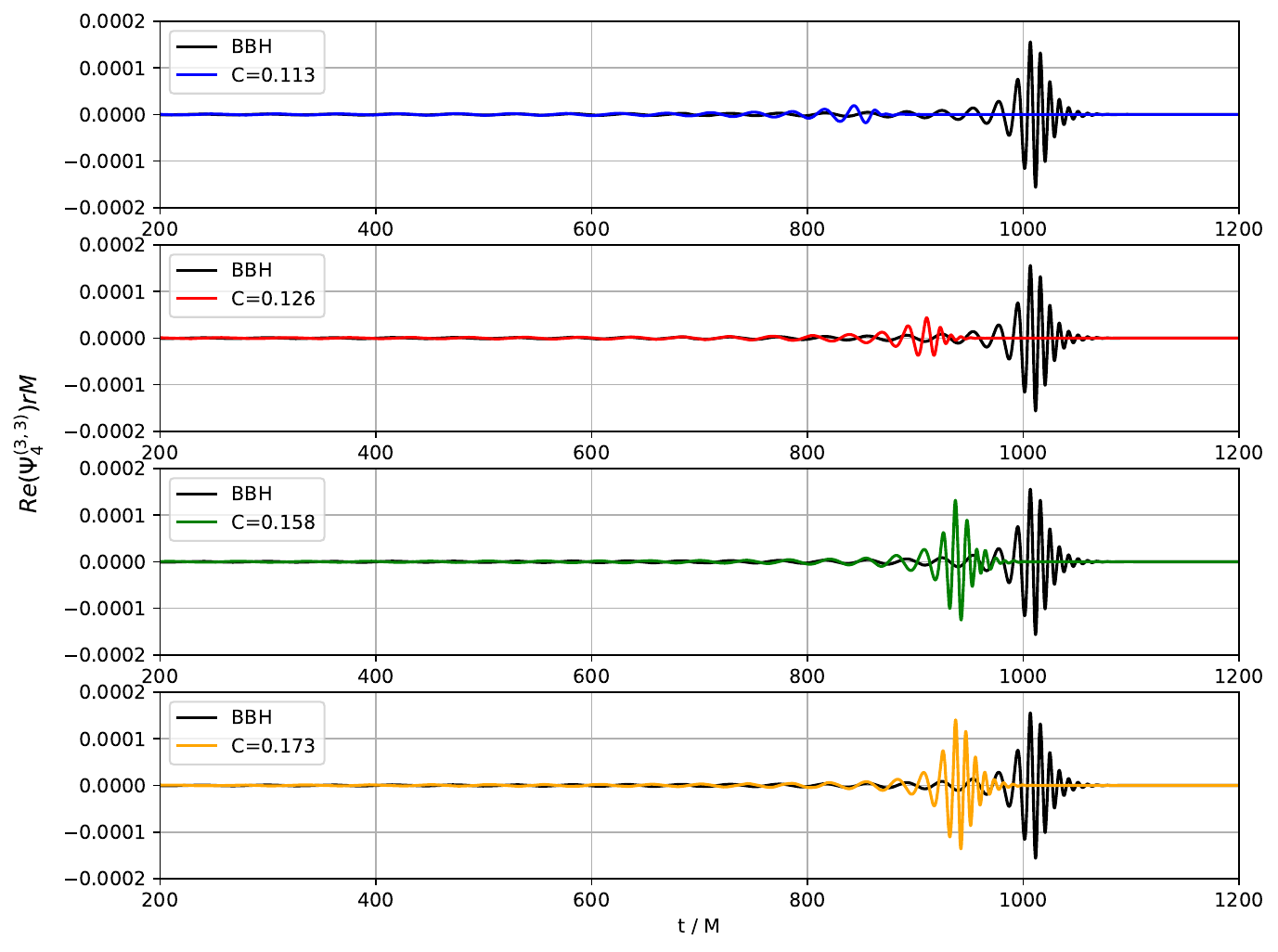}
	\caption{Same as in Fig.~\ref{fig:psi4_22_q5} but the mode $Re(\Psi_4^{(3,3)})$.}
\label{fig:psi4_33_q5}
\end{figure}

Next, we discuss the \gw{} signatures. Figures~\ref{fig:psi4_22_q3} and~\ref{fig:psi4_33_q3} show respectively comparisons of the real part of the (2,2) and (3,3) modes of the $\Psi_4$ Weyl scalar between \bbh{} and \nsbh{} simulations. Panels from top to bottom are arranged in order of increasing compactness. The figures only show the waveforms after the passing of the junk radiation. It is evident from the retarded time $T_{mx}$ when the amplitude of $\Psi_4$ reaches its maximum that \nsbh{} binaries merge earlier than the corresponding mass ratio \bbh{s} (for specific values of $T_{mx}$ in the (2,2) mode see Table~\ref{tab:tmax}). This is expected as stellar deformation leads to an additional correction term in the gravitational potential, which accelerates the mergers by larger dissipation of \gw{s} \cite{Shibata:2011jka}. The same comparison with similar characteristics but for the $q=5$ case is depicted in Figs.~\ref{fig:psi4_22_q5} and~\ref{fig:psi4_33_q5}. We observe the trend that merger occurs later as compactness increases. This is expected since, as mentioned before, the \nsbh{} binary behaves closer to a \bbh{} as compactness increases. At the same time, the time to reach peak amplitude $T_{mx}$ is larger since these $q=5$ binaries are not as luminous as the $q=3$. An interesting observed feature is that the monotonic dependence of $T_{mx}$ with compactness $C$ in $q=5$ does not translate into the $q=3$. The \nsbh{} with $C=0.199$ seems to merge earlier than the $C=0.179$ (see Table~\ref{tab:c5_table_merger}). We have not found a reason for this behavior.

\begin{figure}[!htbp]
    \centering
    \begin{subfigure}{\textwidth}
        \includegraphics[width=\linewidth]{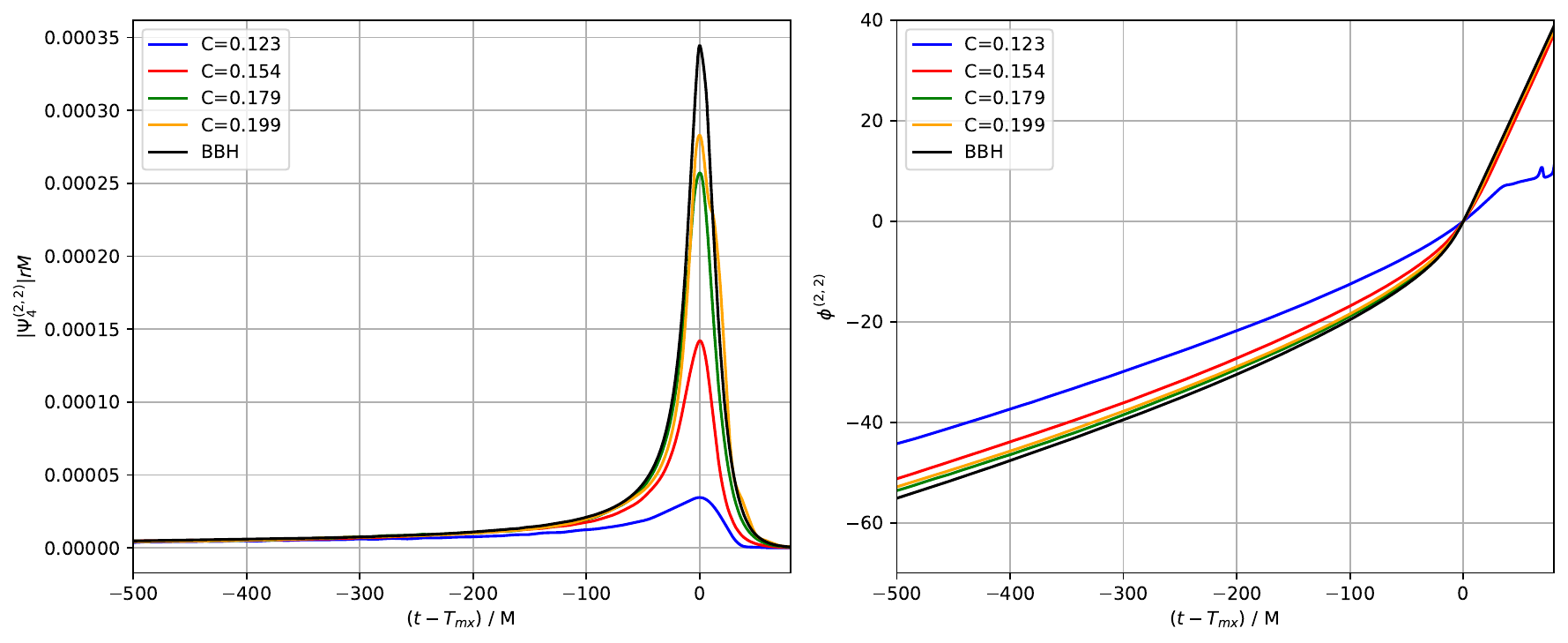}
        \caption{$q=3$}
    \end{subfigure}
    \\
    \centering
    \begin{subfigure}{\textwidth}
        \includegraphics[width=\linewidth]{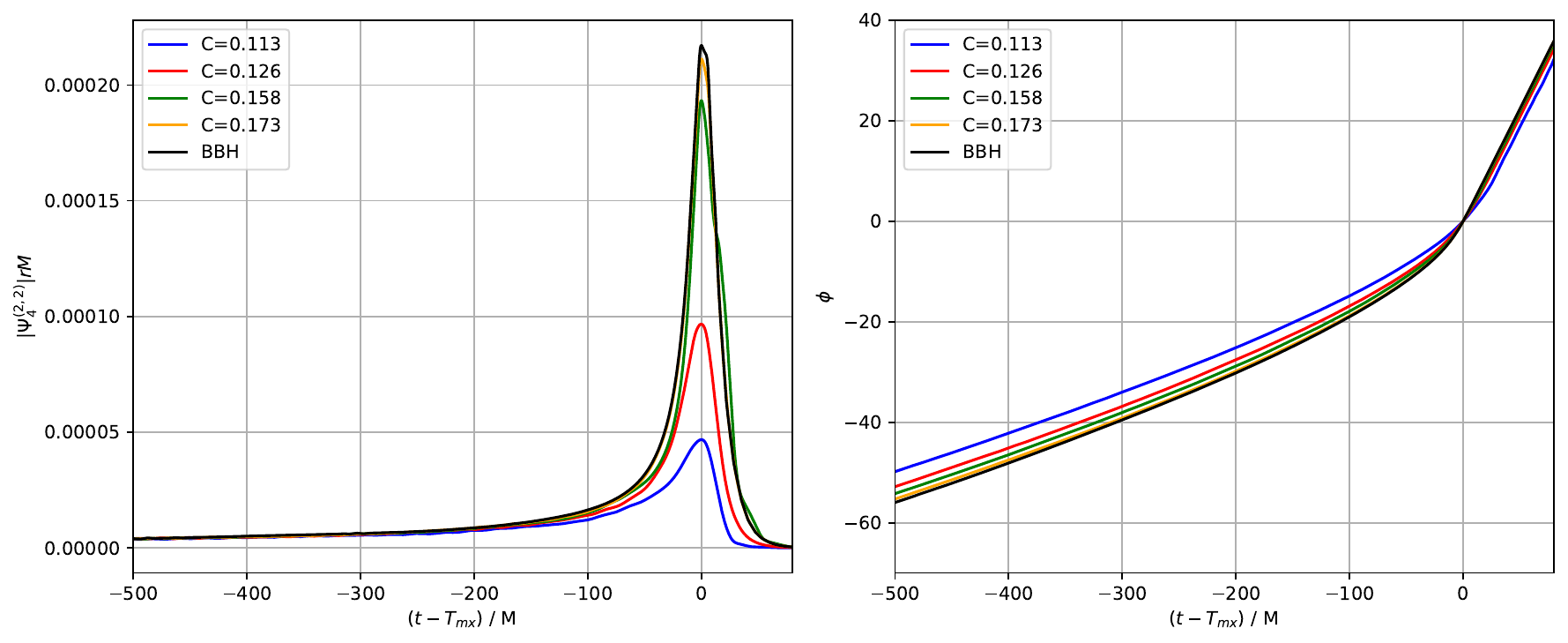}
        \caption{$q=5$}
    \end{subfigure}
    \caption{Amplitude (left panels) and phase (right panels) of the $(2,2)$ mode of $\Psi_4$ for the $q=3$ (top panels) and $q=5$ (bottom panels). Amplitude comparison between \bbh{} and \nsbh{} configurations. The amplitudes have been shifted in time to agree at peak. The same time shift was applied to the phase plus a phase shift, so the phase align at peak amplitude.}
    \label{fig:ch5_psi4amp}
\end{figure}

Another way of analyzing the differences in the waveforms is by looking separately at the amplitude and phase for the (2,2) mode of $\Psi_4$. Figure~\ref{fig:ch5_psi4amp} depicts the amplitude (left panels) and phase (right panels) of $\Psi_4$ for $q=3$ (top panels) and $q=5$ (bottom panels) cases. The amplitudes have been shifted in time to agree at peak. The time shifts are the merger times given in Table~\ref{tab:tmax} added with $130\,M$ for radiation to propagate. The same time shift was applied to the phase plus a phase shift, so the phases align at peak amplitude. The amplitude of $\Psi_4$ increases monotonically with compactness but remains smaller than in the \bbh{} case. This is consistent with \nsbh{} binaries radiating less energy and angular momentum, as we shall see in Sec~\ref{sec:results-finalBH} in Table~\ref{tab:c5_table_merger}.

\begin{table}[!htbp]
    \centering
    \begin{tabular}{ l l l l }
     \toprule
     $q =3$ & & $q=5$ &\\
     \midrule
     \tabhead{$C$} & \tabhead{$T_{mx}/M$} & \tabhead{$C$} & \tabhead{$T_{mx}/M$} \\ 
          \midrule
 BBH   &  758 & BBH   & 877\\
 0.123 &  497 &  0.113 & 727  \\
 0.154 &  664 & 0.126 & 785 \\
 0.179 &  705 & 0.158 & 804 \\
 0.199 &  683 & 0.173 & 807 \\
     \bottomrule
    \end{tabular}
         \caption{ $T_{mx}$ is the retarded time at which the amplitude of the (2,2) mode of $\Psi_4$ reaches a maximum with an extraction radius of $130\,M$.}
    \label{tab:tmax}
\end{table}

\subsection{Spectra and Mismatches}

\begin{figure}[!htbp]
\centering
    \begin{subfigure}{0.49\textwidth}
        \centering
        \includegraphics[width=0.98\linewidth]{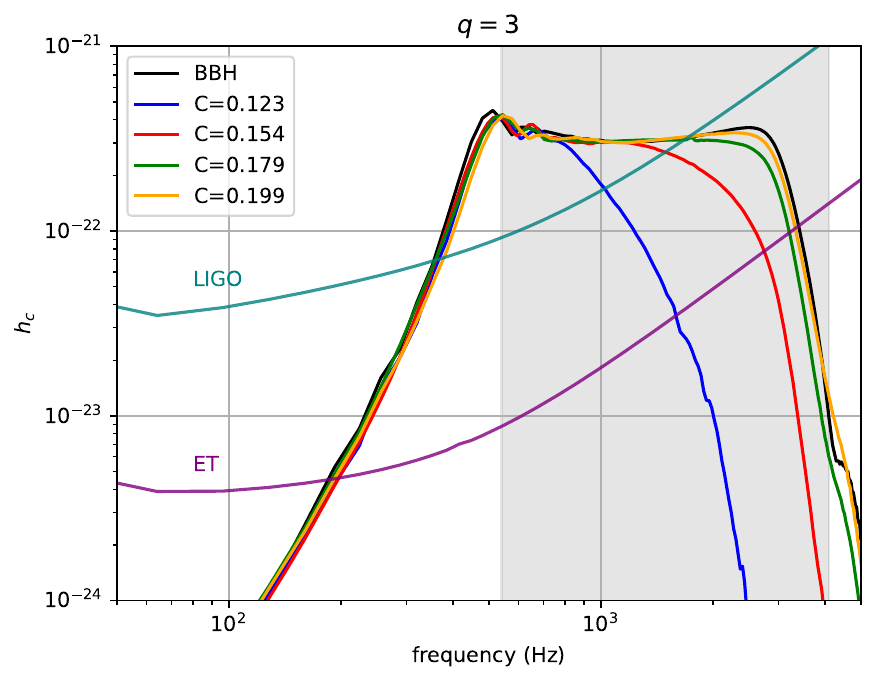}
        
    \end{subfigure}
    \begin{subfigure}{0.49\textwidth}
        \centering
        \includegraphics[width=0.98\linewidth]{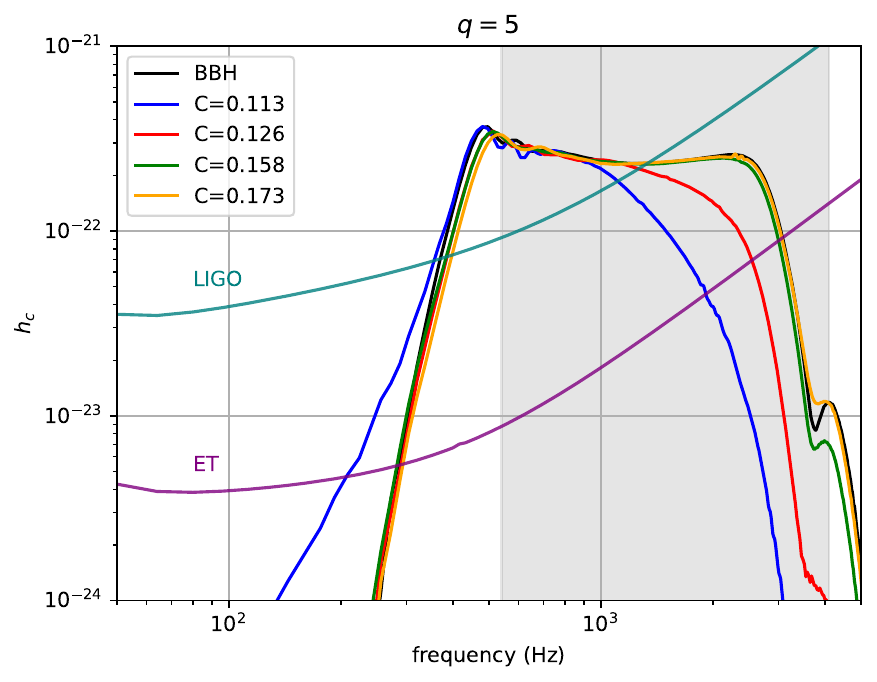}
    \end{subfigure}   
    \caption{Fourier spectrum of the characteristic strain $h_c$ for mass ratio $q=3$ (left) and $q=5$ (right), on top of the sensitivity curves for LIGO and ET.}
\label{fig:ch5_bh_strain_fft}
\end{figure}

It is important to investigate also the differences and similarities in the waveforms with the framework commonly used in \gw{} data analysis. We follow the procedures in Ref.~\cite{Moore_2014} and use the data analysis tool from LSC Algorithm Library~\cite{lalsuite_gitlab}. First, we compute the spectrum of the characteristic strain $h_c(f) = 2 f |\tilde{h}(f)|$,
with
$
    |\tilde{h}(f)|^2 = (|\tilde{h}_+(f)|^2+|\tilde{h}_\times(f)|^2)/2\,.
$
Here, $\tilde{h}_+(f)$ and $\tilde{h}_\times(f)$ are the Fourier transform of the plus and cross-polarization of the strain at a distance of $150$ Mpc, calculated from $\Psi_4$ including up to the $l = 8$ mode. Figure~\ref{fig:ch5_bh_strain_fft} shows the spectrum of $h_c$ for all compactness cases under consideration and for a \bbh{} (left panel for $q=3$ and right panel for $q=5$). The figures include the sensitivity curves for LIGO (obtained from LSC Algorithm Library for second-generation detectors) and the Einstein Telescope. Not surprisingly, the larger the compactness, the closer the spectrum for $h_c$ is to the \bbh{} case. 

Next, we calculate mismatches between the strains from \nsbh{} binaries and the one from a \bbh{.} The mismatches are computed from  $1 - \mathcal{O}(h_1 | h_2)$, with the match or overlap given by
\begin{equation}
    \mathcal{O}(h_1 | h_2) = \frac{\max_{\phi_c,t_c}\langle h_1 | h_2\rangle}{\sqrt{\langle h_1 | h_1\rangle\langle h_2 | h_2\rangle}}
\end{equation}
where
\begin{equation}
    \langle h_1 | h_2 \rangle \equiv 4 \textrm{Re}\int_{f_{mn}}^{f_{mx}}{\frac{\tilde{h}_1^*(f) \tilde{h}_2(f)}{S_n(f)}}df\,.
    \label{eq:match}
\end{equation}
The maximization is over coalescence time $t_c$ and coalescence phase $\phi_c$. In Eq.~(\ref{eq:match}), $S_n$ is the one-sided power spectral density (sensitivity curve) of the detector noise. We considered sensitivity curves for LIGO and the Einstein Telescope (ET). 

Table~\ref{tab:c5_table_mismatch} shows the mismatches of the plus polarization $h_+$ for $q=3$ and 5 at inclination angles $\iota=0, \pi/6$, and $\pi/3$. To eliminate errors due to Gibbs phenomena when taking the FFTs, we taper the first $3$ cycles of the waveform and start the integration in Eq.~(\ref{eq:match}) at $f_{mn}$. The upper limit of integration is set to $f_{mx}$ to cover the merger as shown in Fig.\ref{fig:ch5_bh_strain_fft} in the shaded area. The integration range for $q=3$ is $[f_{mn},f_{mx}]=[320,4096]$Hz, and for $q=5$, $[f_{mn},f_{mx}]=[550,4096]$Hz. These bounds of integration are set to ensure a fair comparison between mismatches across cases with different compactness, detectors, and inclination angles.

\begin{table}[!htbp]  
    \centering
    \begin{tabular}{l | l  l | l l | l l }
     \toprule
               &        \multicolumn{6}{c}{$q =3$} \\ 
\midrule
   & \multicolumn{2}{c}{$\iota=0$} & \multicolumn{2}{c}{$\iota=\pi/6$} & \multicolumn{2}{c}{$\iota=\pi/3$} \\
    \midrule
     $C$  & LIGO   & ET     & LIGO   & ET     & LIGO   & ET      \\
    0.123  & 0.0796 & 0.0844 & 0.1071 & 0.1116 & 0.1490 & 0.1537  \\
    0.154  & 0.0123 & 0.0125 & 0.0371 & 0.0376 & 0.0845 & 0.0856  \\
    0.179  & 0.0080 & 0.0080 & 0.0371 & 0.0377 & 0.0922 & 0.0940  \\
    0.199  & 0.0104 & 0.0102 & 0.0346 & 0.0347 & 0.0739 & 0.0746  \\
         \midrule
          &        \multicolumn{6}{c}{$q =5$} \\ 
          \midrule
   & \multicolumn{2}{c}{$\iota=0$} & \multicolumn{2}{c}{$\iota=\pi/6$} & \multicolumn{2}{c}{$\iota=\pi/3$} \\
          \midrule
    $C$    & LIGO   & ET     & LIGO   & ET     & LIGO   & ET      \\
    0.113  & 0.0350 & 0.0391 & 0.0667 & 0.0739 & 0.1239 & 0.1372  \\
    0.126  & 0.0107 & 0.0115 & 0.0431 & 0.0480 & 0.1094 & 0.1209  \\
    0.158  & 0.0040 & 0.0040 & 0.0326 & 0.0361 & 0.0957 & 0.1061  \\
    0.173  & 0.0076 & 0.0073 & 0.0073 & 0.0074 & 0.0118 & 0.0120  \\
     \bottomrule
    \end{tabular}
     \caption{Mismatches between the \nsbh{} waveforms and the waveform of a \bbh{} for three inclination angle $\iota$ and sensitivity curves for LIGO and the Einstein Telescope.}
         \label{tab:c5_table_mismatch}
\end{table}

For a given mass ratio and inclination, the mismatches decrease with increasing compactness since, as mentioned before, an increase in compactness brings the \nsbh{} binary to look more like a \bbh{}. The exception is the $q=3$, $C=0.199$ case. Its mismatches are larger than for $q=3$, $C=0.179$. This is connected to the observation made in Sec.~\ref{sec:results-GW} that the $q=3$, $C=0.199$ \nsbh{} binary merges earlier than the $q=3$, $C=0.179$ binary.  Finally, as the inclination angle increases, the mismatch increases accordingly. For $\iota=\pi/3$, mismatches for mass ratio $q=3$ can go up $0.07-0.15$ while $q=5$ mismatches go up to $0.01-0.14$. This indicates that the inclusion of higher-order modes reveals more information to distinguish between \nsbh{} and \bbh{} waveforms~\cite{Khamesra_2021}. The standard mismatch cutoff for LIGO sensitivity to claim two waveforms are similar enough is $\lesssim 0.02$. Thus, from the results in Table~\ref{tab:c5_table_mismatch}, only the $C=0.173$, $q=5$ is ``indistinguishable" from a \bbh{} independently of the inclination angle. This is consistent with what we have been pointing out several times that this case is the one that behaves closer to a \bbh{}. If we focus on zero inclination angle, some cases with lower compactness have mismatches $< 0.02$, thus also ``indistinguishable." 

\subsection{Impact of the spurious neutron star oscillation}
\label{sec:results-osc}

\begin{figure}[!htbp]
    \centering
    \begin{subfigure}{0.8\textwidth}
        \includegraphics[width=\linewidth]{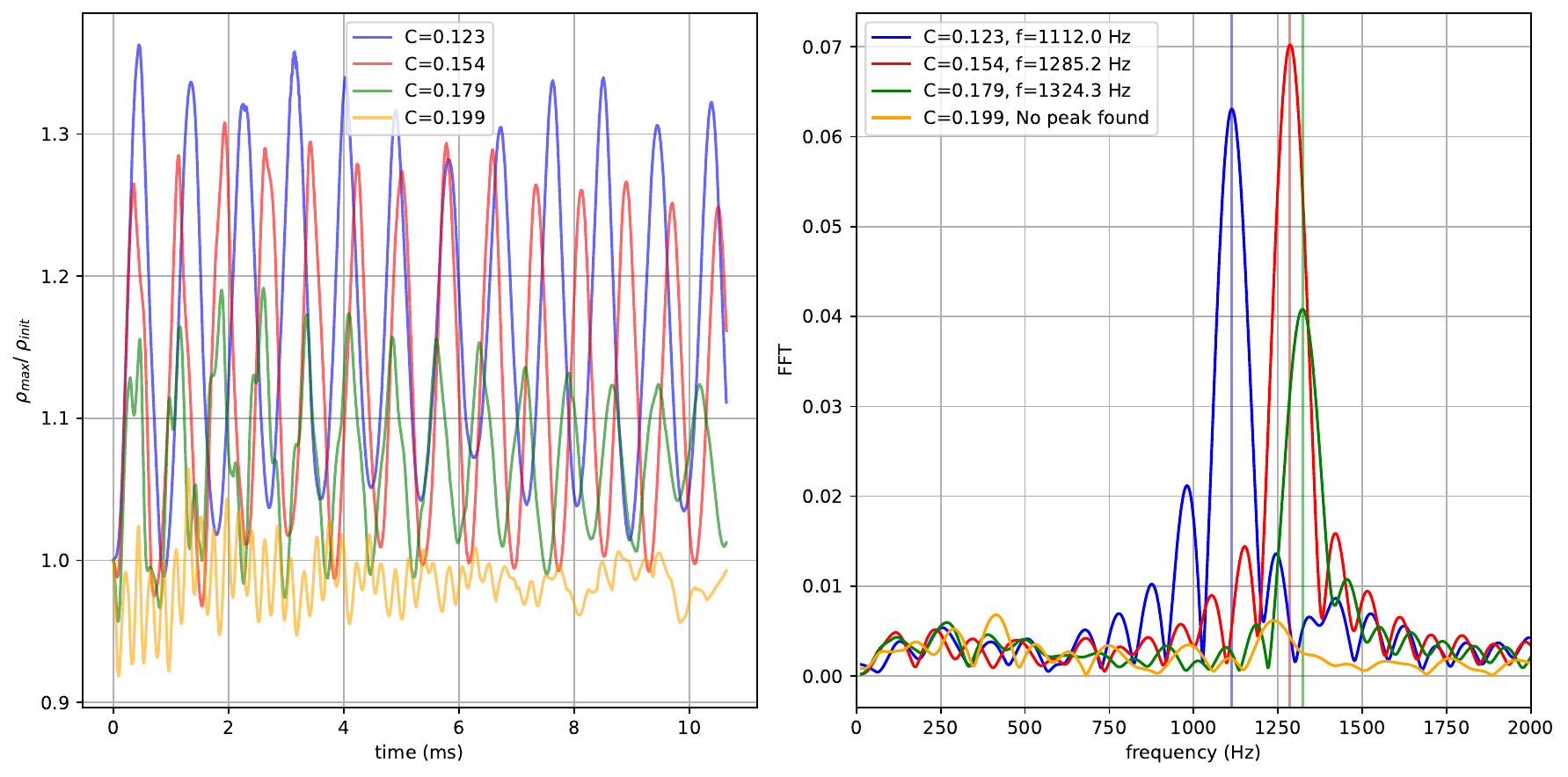}
        \caption{$q=3$}
    \end{subfigure}
    \\
    \centering
    \begin{subfigure}{0.8\textwidth}
        \includegraphics[width=\linewidth]{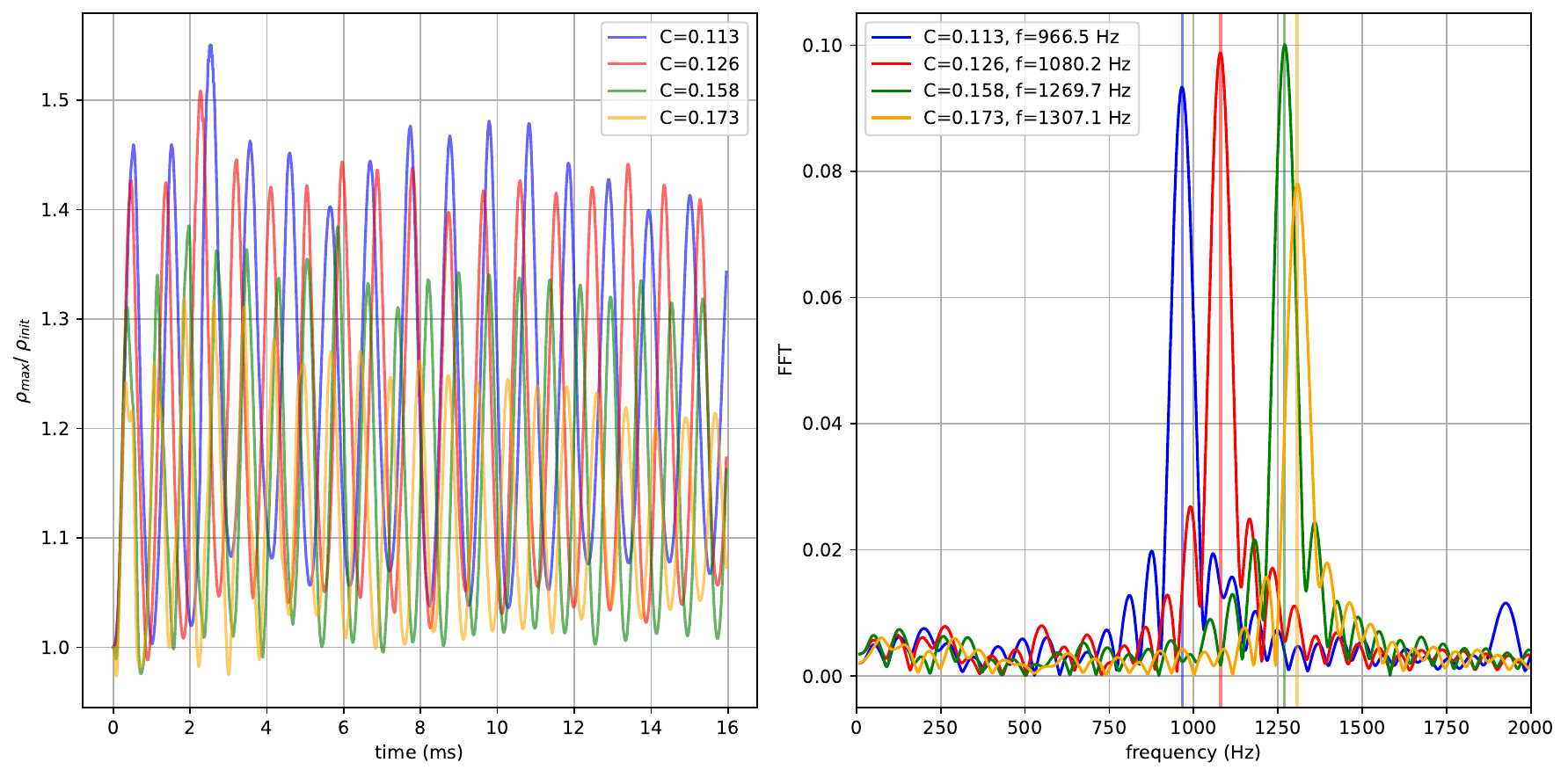}
        \caption{$q=5$}
    \end{subfigure}
    \caption{The left panels show the oscillations in the central density of the \ns{} normalized to the initial central density of the star. The right panels show the corresponding FFT, with the insert showing the peak frequency. No significant peak was found for the case $C=0.199$, $q=3$.}
        \label{fig:ch5_densosc}
\end{figure}

As pointed out in our previous study~\cite{Khamesra_2021}, our framework to construct initial data triggers spurious breathing modes in the \ns{}. Figure~\ref{fig:ch5_densosc} shows the oscillations in the central density of the star due to the ``imperfections" in the initial data. The left panels show the oscillations relative to the initial central density, and the corresponding right panels show the Fourier transforms. Two interesting observations can be made here. First, smaller stars (higher compactness) exhibit smaller amplitude oscillations, while larger stars (lower compactness) can have oscillations in the central density, reaching amplitudes of $\sim 25\%$ of its initial value. The observed oscillations are a manifestation of radial oscillation in the \ns{}~\cite{2001A&A...366..565K} triggered by the initial data setup. Since they are radial oscillations, they do not leave an imprint in the \gw{s} emitted.

\subsection{The Final State}
\label{sec:results-finalBH}

\begin{figure}[!htbp]
\centering
    \begin{subfigure}{0.49\textwidth}
        \centering
        \includegraphics[width=0.98\linewidth]{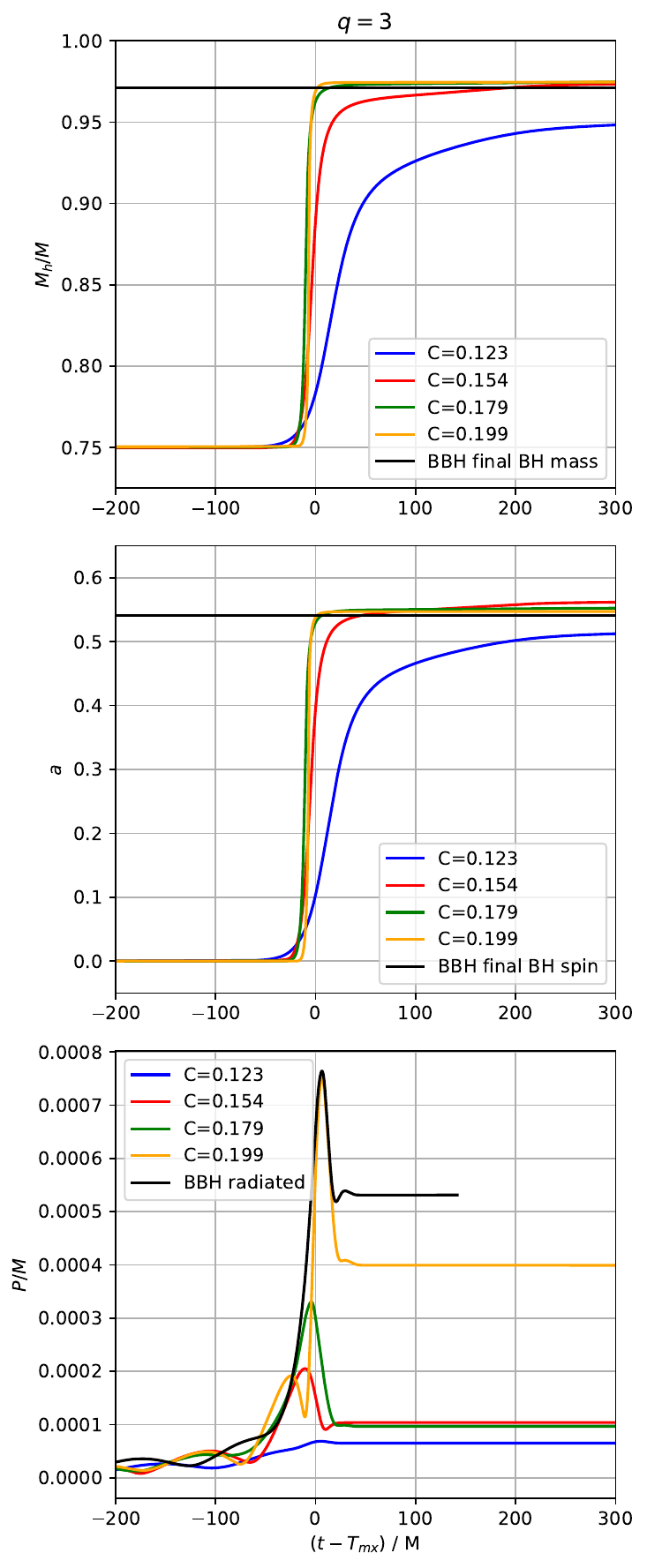}        
    \end{subfigure}
    \begin{subfigure}{0.49\textwidth}
        \centering
        \includegraphics[width=0.98\linewidth]{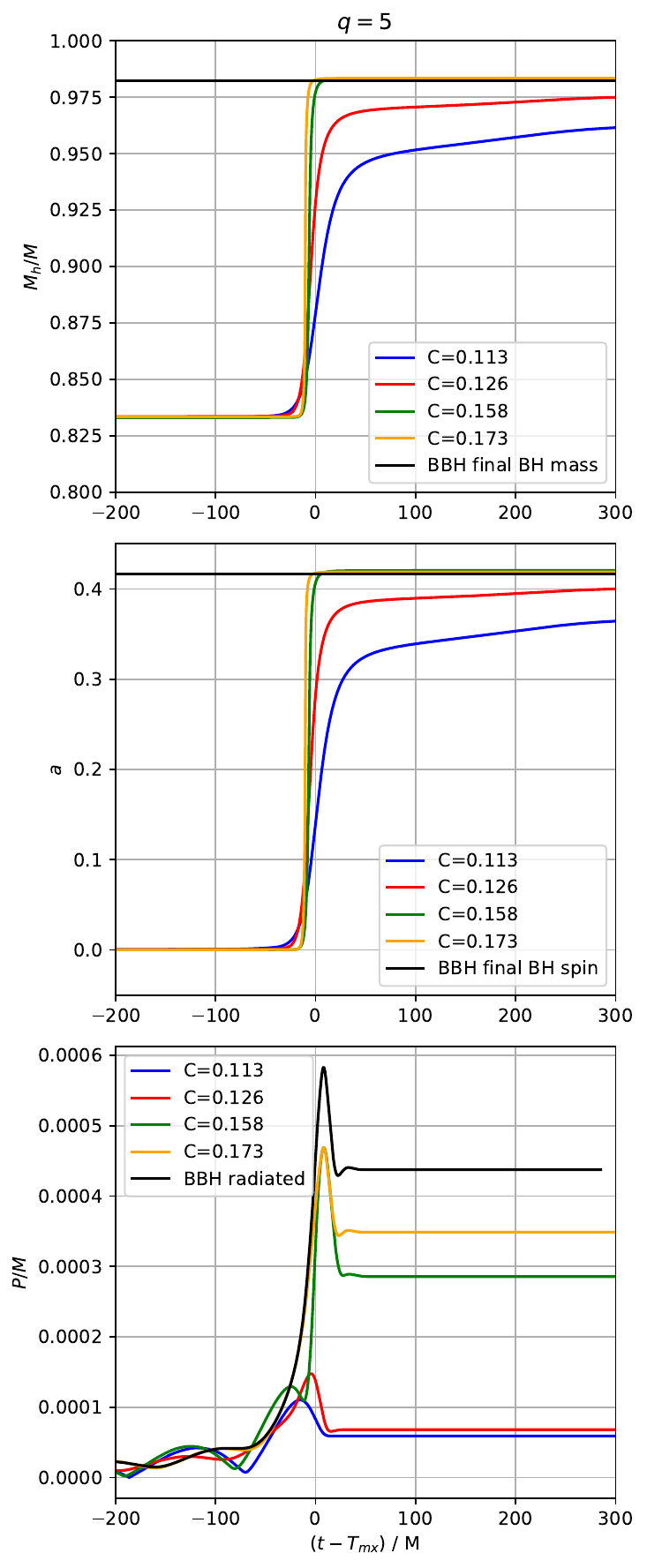}
    \end{subfigure}    
    \caption{ From top to bottom, the mass and spin of the \bh{} and the radiated linear momentum during the course of the simulation. The left panels are for the $q=3$ and the right for $q=5$. For the mass and spin, the black lines denote the value of the mass and spin for the final \bh{} in the \bbh{} merger. The black lines in the bottom panels are the evolution of the radiated linear momentum for the \bbh{} system. Time for each case has been shifted to align when $\Psi_4^{(2,2)}$ reaches peak amplitude. }  
\label{fig:ch5_bh_qlprop}
\end{figure}

Figure~\ref{fig:ch5_bh_qlprop} shows, from top to bottom, the mass and spin of the \bh{} and the radiated linear momentum during the course of the simulation. The left panels are for the $q=3$ and the right for $q=5$. For the mass and spin, the black lines denote the value of the mass and spin for the final \bh{} in the \bbh{} merger. The black lines in the bottom panels are the evolution of the radiated linear momentum for the \bbh{} system. Time for each case has been shifted to align when $\Psi_4^{(2,2)}$ reaches peak amplitude.
The mass and spin of the \bh{} after the merger increase with the compactness of the star. The lower the compactness of the \ns{}, the slower the growth of the mass and spin of final \bh{}. This is because the \bh{} takes longer to accrete the debris from the disrupted \ns{}. In the extreme cases where the \ns{s} have the lowest compactness, $C=0.123$ and 0.113, the mass and spin of the final \bh{} are significantly lower than the corresponding mass and spin of the final \bh{} from a \bbh{} merger. In these cases, between $15\%$ and $10\%$ of the mass and angular momentum of the \ns{} remains trapped in the accretion disk surrounding the \bh{} still present at the end of the simulation.

\begin{table}[!htbp]
    \centering
    \begin{tabular}{ l l l l l l l l r}
     \toprule
     \tabhead{$C$} & \tabhead{$M_T/M$} & \tabhead{$M_h/M$} & \tabhead{$M_{ej}/M$} & \tabhead{$E_{rad}/M$}&  
      \tabhead{$J_T/M^2$} & \tabhead{$a$} & \tabhead{$J_{rad}/M^2$} & \tabhead{$v_k(km/s)$} \\
     \midrule
           \midrule
          &        \multicolumn{6}{c}{$q =3$} \\ 
          \midrule
BBH   & 0.9918 & 0.9713 & 0.00     & 0.0205 & 0.7025 & 0.5405 & 0.1926 & 163.93 \\
0.123 & 0.9644 & 0.9515 & 8.49e-03 & 0.0044 & 0.5531 & 0.5182 & 0.0839 & 20.48 \\
0.154 & 0.9854 & 0.9736 & 1.78e-03 & 0.0100 & 0.6670 & 0.5620 & 0.1343 & 31.95 \\
0.179 & 0.9915 & 0.9747 & 1.33e-03 & 0.0155 & 0.6882 & 0.5523 & 0.1634 & 29.59 \\
0.199 & 0.9926 & 0.9746 & 1.55e-05 & 0.0180 & 0.6921 & 0.5470 & 0.1724 & 122.76 \\
      \midrule
          &        \multicolumn{6}{c}{$q =5$} \\ 
          \midrule
BBH   & 0.9940 & 0.9824 & 0.00     & 0.0116 & 0.5231 & 0.4165 & 0.1212 & 133.54 \\
0.113 & 0.9728 & 0.9624 & 6.16e-03 & 0.0042 & 0.4134 & 0.3665 & 0.0739 & 18.41 \\
0.126 & 0.9888 & 0.9754 & 7.17e-03 & 0.0062 & 0.4698 & 0.4011 & 0.0881 & 20.91 \\
0.158 & 0.9936 & 0.9834 & 2.11e-05 & 0.0102 & 0.5171 & 0.4206 & 0.1103 & 87.07 \\
0.173 & 0.9944 & 0.9834 & 1.48e-04 & 0.0108 & 0.5159 & 0.4189 & 0.1107 & 106.31 \\
     \bottomrule
    \end{tabular}
         \caption{Final properties of the mergers. $M_T = M_h + M_{ej} + E_{rad}$ with $M_h$ the mass of the final \bh{} computed from the Christodoulou mass, $M_{ej}$ the mass ejected beyond a radius $35\,M$ from the binary, and $E_{rad}$ the energy radiated in \gw{s}. $J_T/M^2 = a (M_h/M)^2 + J_{rad}/M^2$, with $a$ the dimensionless spin of the final \bh{} and $J_{rad}$ the angular momentum radiated in \gw{s}. $v_k$ is the kick of the final  \bh{}. For reference, $E_{ADM}/M =  0.9917 \,(q=3)$, $E_{ADM}/M = 0.994 \, (q=5)$, $J_{ADM}/M^2 = 0.7023\, (q=3)$, and  $J_{ADM}/M^2 = 0.523\, (q=5)$ at the start of the simulation.}
    \label{tab:c5_table_merger}
\end{table}

Table \ref{tab:c5_table_merger} shows the final properties of the mergers. $M_T = M_h + M_{ej} + E_{rad}$ is the total mass-energy at the end of the simulation, with $M_h$ the mass of the final \bh{} computed from the Christodoulou mass, $M_{ej}$ the mass ejected beyond a radius $35\,M$ from the binary, and $E_{rad}$ the energy radiated in \gw{s}. $J_T/M^2 = a(M_h/M)^2 + J_{rad}/M^2$ is the total angular momentum with $a$ the dimensionless spin of the final \bh{} and $J_{rad}$ the angular momentum radiated in \gw{s}. $v_k$ is the kick of the final  \bh{}. For reference, the initial energy and angular momentum are $E_{ADM}/M =  0.9917 \,(q=3)$, $E_{ADM}/M = 0.994 \, (q=5)$, $J_{ADM}/M^2 = 0.7023\, (q=3)$, and  $J_{ADM}/M^2 = 0.523\, (q=5)$ at the start of the simulation. 

In the highest compactness cases, $M_T/M$ exceeds $M_{ADM}/M$. This is because of the inaccuracy of calculating $E_{rad}/M$ at a finite distance from the binary. 
On the other hand, for low compactness, $J_T/M^2$ is significantly smaller than $J_{ADM}/M^2$ because of the material that is not accounted for within the radius where the ejected material is calculated.

From $E_{rad}$ and $J_{rad}$ in Table \ref{tab:c5_table_merger}, we observe that the \bbh{} system is the one that produced the largest radiation. In contrast, the \nsbh{} systems produce smaller radiation emission, but it grows monotonically with the compactness. There are two main reasons for this. Low compactness yields disruption, which damps \gw{} emission. The other reason is, as mentioned before, low compactness accelerates the merger and, thus, restraints the total amount of emitted radiation.

\begin{figure}[!htbp]
\centering
\includegraphics[width=\textwidth]{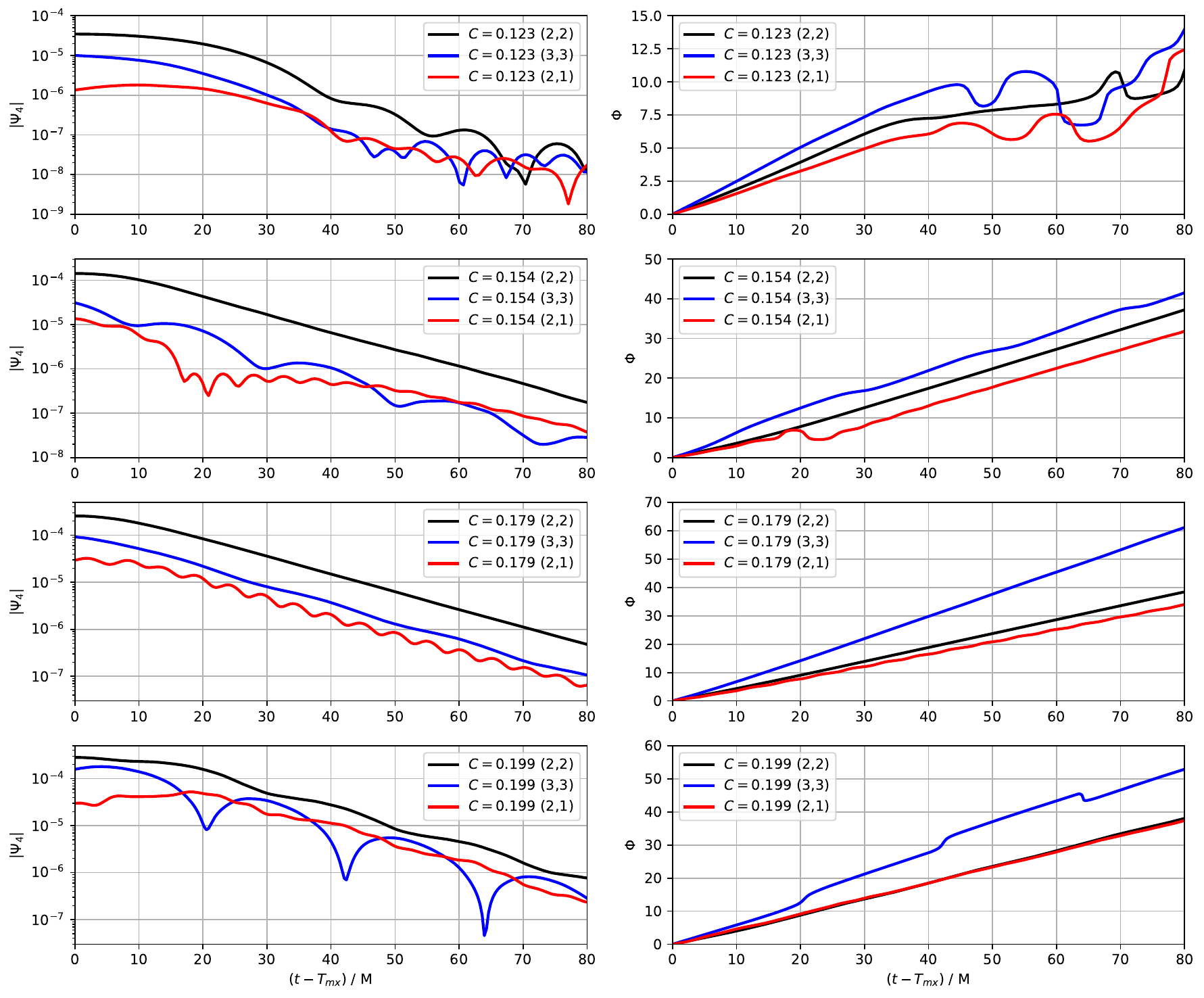}
	\caption{Post-merger $\Psi_4$ signal for $q=3$.  The left panels depict the amplitude of $\Psi_4$ for the modes (2,1), (2,2), and (3,3), with the right panels showing their phases. The times have been shifted so amplitude and phase align at peak luminosity.} 
\label{fig:ch5_qnmPlots_q3}
\end{figure}

\begin{figure}[!htbp]
\centering
\includegraphics[width=\textwidth]{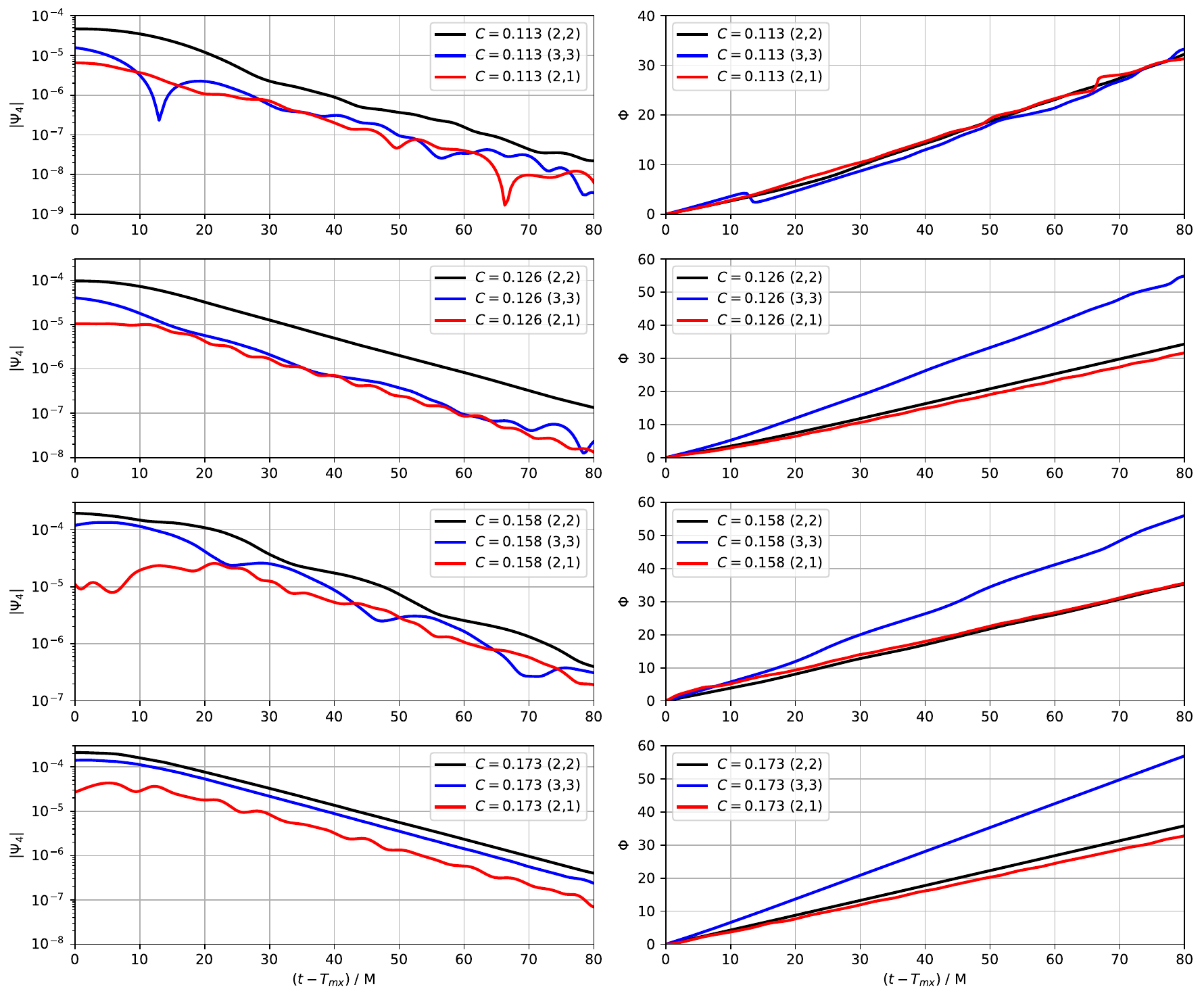}
	\caption{Same as in Fig.~\ref{fig:ch5_qnmPlots_q3} but for $q=5$.}
\label{fig:ch5_qnmPlots_q5}
\end{figure}

\subsection{Quasi-normal ringing}
\label{sec:results-QNM}

Figures~\ref{fig:ch5_qnmPlots_q3} and \ref{fig:ch5_qnmPlots_q5} show the post-merger $\Psi_4$ signal for $q=3$ and 5, respectively.  The left panels depict the amplitude of $\Psi_4$ for the modes (2,1), (2,2), and (3,3), with the right panels showing their phases. The times have been shifted so amplitude and phase align at peak luminosity. Given the masses and spins of the final \bh{} from Table~\ref{tab:c5_table_merger}, the top half in  Tables~\ref{tab:c5_table_qnm_q3} and \ref{tab:c5_table_qnm_q5} show the \QNM{} frequencies $\omega$ and damping times $\tau$ for each mode from Ref.~\cite{Berti:2005ys}. The bottom half of these tables shows the \QNM{} values for $\omega$ and $\tau$ from fittings of $\Psi_4 \propto \exp{(-t/\tau)}\sin{(\omega\,t)}$ to our simulation data. For low compactness, the signal does not exhibit a {\it clean} \QNM{} ringing. This is because the \bh{} is accreting tidal debris. The oscillations observed in the amplitude for high compactness are due to mode mixing in our extraction procedure. 

\begin{table}[!htbp]
    \centering
    \begin{tabular}{l l l l l l l }
     \toprule
     \tabhead{$C$} & \tabhead{$\omega_{(2,1)}$} & \tabhead{$\tau_{(2,1)}$} & \tabhead{$\omega_{(2,2)}$} & \tabhead{$\tau_{(2,2)}$} & \tabhead{$\omega_{(3,3)}$} & \tabhead{$\tau_{(3,3)}$} \\
     \midrule
           \midrule
          &        \multicolumn{5}{c}{From Ref.~\cite{Berti:2005ys}} \\ 
          \midrule
     0.123 & 0.445 & 11.639 & 0.493 & 11.726 & 0.787 & 11.326 \\
     0.154 & 0.442 & 11.738 & 0.496 & 11.839 & 0.790 & 11.462 \\
     0.179 & 0.440 & 11.716 & 0.492 & 11.814 & 0.784 & 11.431 \\
     0.199 & 0.439 & 11.704 & 0.491 & 11.800 & 0.782 & 11.415 \\
                \midrule
          &        \multicolumn{5}{c}{From Simulations} \\ 
          \midrule
     0.123 & 0.080 & 10.622 & 0.084 &  8.112 & 0.077 & 10.923 \\
     0.154 & 0.472 & 22.001 & 0.491 & 11.085 & 0.488 & 10.838 \\
     0.179 & 0.434 & 11.682 & 0.491 & 11.534 & 0.783 & 11.181 \\
     0.199 & 0.471 & 11.288 & 0.489 & 11.301 & 0.653 & 12.045 \\

     \bottomrule
    \end{tabular}
         \caption{\QNM{s} frequency $\omega$ and decaying time scale $\tau$ in units of the final \bh{} mass $M_h$ for the most dominant modes for the $q=3$ cases. The top half are values from Ref.~\cite{Berti:2005ys} and the bottom half from fittings to our simulation data.}
    \label{tab:c5_table_qnm_q3}
\end{table}

\begin{table}[!htbp]
    \centering
    \begin{tabular}{l l l l l l l }
     \toprule
     \tabhead{$C$} & \tabhead{$\omega_{(2,1)}$} & \tabhead{$\tau_{(2,1)}$} & \tabhead{$\omega_{(2,2)}$} & \tabhead{$\tau_{(2,2)}$} & \tabhead{$\omega_{(3,3)}$} & \tabhead{$\tau_{(3,3)}$} \\
     \midrule
          \midrule
          &        \multicolumn{5}{c}{From Ref.~\cite{Berti:2005ys}} \\ 
          \midrule
     0.113 & 0.420 & 11.424 & 0.450 & 11.472 & 0.720 & 11.042 \\
     0.126 & 0.418 & 11.456 & 0.451 & 11.509 & 0.722 & 11.089 \\
     0.158 & 0.418 & 11.485 & 0.452 & 11.544 & 0.723 & 11.125 \\
     0.173 & 0.417 & 11.482 & 0.452 & 11.541 & 0.723 & 11.122 \\
              \midrule
          &        \multicolumn{5}{c}{From Simulations} \\ 
          \midrule
     0.113 & 0.429 & 10.451 & 0.444 & 10.091 & 0.459 & 10.612 \\
     0.126 & 0.420 & 10.410 & 0.449 & 10.954 & 0.718 & 10.705 \\
     0.158 & 0.434 & 12.312 & 0.452 & 11.120 & 0.724 & 10.962 \\
     0.173 & 0.417 & 11.231 & 0.451 & 11.345 & 0.722 & 11.012 \\
     \bottomrule
    \end{tabular}
         \caption{Same as in Table~\ref{tab:c5_table_qnm_q3} but for the $q=5$.}
    \label{tab:c5_table_qnm_q5}
\end{table}

\section{Conclusions}
\label{sec:conclusions}

The present study  investigated the effect of varying the compactness of the \ns{} on \nsbh{} mergers with polytropic \eos{.} We considered four compactness cases for mass ratios $q=3$ and $5$ to compare the inspiral and merger dynamics,  the \gw{s} emitted, the properties of the final \bh{}, including its \QNM{} ringing.
The \nsbh{} system behaves more and more like a \bbh{} as the compactness increases. For low compactness, the tidal debris resembles an accretion, and the \gw{} signal is dramatically different from that of a \bbh{} with the same mass ratio. The compactness also affects the merger time. For low compactness, the binary merges $100 - 200\,M$ earlier than then \bbh{}. The \QNM{} ringing of the final \bh{} was significantly affected by the effects of tidal disruption. Clean \QNM{} modes are only observed for high compactness cases. For low compactness, the tidal debris accreting into the final \bh{} contaminates the signal. Mismatches between \nsbh{} and \bbh{} waveforms showed that the mismatches decrease for higher compactness cases, while the inclusion of higher-order modes for larger inclination angles leads to higher mismatches. When put together, the \QNM{} and mismatches results, \nsbh{} with low compactness are the best candidates to investigate the signatures from \nsbh{} coalescences. 

\paragraph*{Acknowledgements}
This work is supported by NSF grants PHY-2114582 and PHY-2207780. We would like to thank Kostas Kokkotas, Deborah Ferguson, Deirdre Shoemaker, Aasim Jan, and Francois Foucart for insightful discussions and for sharing their resources for this work. 

\vspace{0.25in}
\bibliographystyle{iopart-num}
\bibliography{refs}
\end{document}